\documentclass[12pt,superscriptaddress,reprint]{article}
\usepackage{amsmath,amssymb}
\usepackage{lineno}
\usepackage{color}
\usepackage{graphicx}
\usepackage{cite}
\usepackage[utf8]{inputenc}
\usepackage[T1]{fontenc}
\usepackage{authblk}
\newcommand{\be}{\begin{equation}} 
\newcommand{\ee}{\end{equation}}
\newcommand{\beq}{\begin{eqnarray}}
\newcommand{\eeq}{\end{eqnarray}}
\newcommand{\ba}{\begin{array}}    
\newcommand{\ea}{\end{array}}
\newcommand{\crs}{cross section}
\newcommand{\crss}{\crs s}
\newcommand{\diff}{differential} 
\newcommand{\photo}{photoproduction}
\newcommand{\fsi}{final-state interaction}

\newcommand{\gN}{\gamma N}
\newcommand{\gn}{\gamma\!N}
\newcommand{\gNpiN}{\gamma N\!\to\pi N}
\newcommand{\gnpiN}{\gamma n\!\to\pi N}

\newcommand{\gppip}{\gamma p\to\pi^0 p}
\newcommand{\gppin}{\gamma p\to\pi^+ n}
\newcommand{\gnpip}{\gamma n\!\to\pi^- p}
\newcommand{\gnpin}{\gamma n\!\to\pi^0 n}

\newcommand{\gd}{\gamma d}
\newcommand{\gdpi}{\gd\to\pi N\!N}         
\newcommand{\bfgdpi}{\mbox {\boldmath $\gdpi$}} 
\newcommand{\gdpipn}{\gd\to\pi^0 pn}         
\newcommand{\gdpinn}{\gd\to\pi^+ nn}         
\newcommand{\gdpipp}{\gd\to\pi^- pp}         
\newcommand{\gdpiNN}{\gd\to\pi N\!N}         
\newcommand{\bfpipp}{\mbox {\boldmath $\gdpipp$}}
\newcommand{\bfpinn}{\mbox {\boldmath $\gdpinn$}}

\newcommand{\dsdo}{d\sigma/d\Omega}
\newcommand{\dso}{d\sigma\!/d\Omega}

\newcommand{\sgn}{\sigma_{\gamma n}}
\newcommand{\sgd}{\sigma_{\gamma d}}
\newcommand{\hM}{\hat M}
\newcommand{\mgn}{M_{\gamma n}}
\newcommand{\hmgn}{{\hM}_{\gamma n}}
\newcommand{\hPsi}{\hat\Psi_d}  
\newcommand{\mgd}{M_{\gamma d}}
\newcommand{\asgn}{\bar{\sigma}_{\gamma n}}
\newcommand{\bfK}{\mbox {\boldmath $K$}}
\newcommand{\bfp}{\mbox {\boldmath $p$}}

\newcommand{\dist}{\displaystyle}
\newcommand{\rpt}{\rule{0pt}{14pt}}
\newcommand{\rptt}{\rule{0pt}{16pt}}
\newcommand{\rttt}{\rule{0pt}{23pt}} 
\newcommand{\rtttt}{\rule{0pt}{25pt}} 

\newcommand{\ega}{E_{\gamma}}
\newcommand{\qga}{q_{\gamma}}

\newcommand{\half}{\frac{1}{2}}

\newcommand{\bfq}{\mbox {\boldmath $q$}}

\newcommand{\bfr}{\mbox {\boldmath $r$}}
\newcommand{\bfk}{\mbox {\boldmath $k$}}
\newcommand{\bfL}{\mbox {\boldmath $L$}}
\newcommand{\bfn}{\mbox {\boldmath $n$}}

\newcommand{\bfe}{\mbox {\boldmath $e$}}

\newcommand{\bfx}{\mbox {\boldmath $x$}}
\newcommand{\bfy}{\mbox {\boldmath $y$}}

\newcommand{\bfsig}{\mbox {\boldmath $\sigma$}}
\newcommand{\bfeps}{\mbox {\boldmath $\epsilon$}}
\newcommand{\bfdeL}{\mbox {\boldmath $\Delta$}}

\newcommand{\hA}{\hat A}

\title{\large\bf On the Photoproduction Reactions $\bfgdpi$}

\date{\today}

\author[1]{William~J.~Briscoe}
\author[2,1]{Alexander~E.~Kudryavtsev}
\author[1]{Igor~I.~Strakovsky\footnote{Corresponding author: \texttt{igor@gwu.edu}}}
\author[2]{Vladimir~E.~Tarasov}
\author[1]{Ron~L.~Workman}
\affil[1]{\small{Institute for Nuclear Studies, Department of Physics, The George Washington University, Washington, DC 20052, USA}}
\affil[2]{National Research Centre ``Kurchatov Institute'', Institute for Theoretical and Experimental Physics (ITEP), Moscow  117218, Russia}

\begin{document}

\maketitle


\vspace{-4mm}
\begin{abstract}
A review of our works  providing a theoretical description of incoherent pion \photo\ on the deuteron is presented. The existing $\gdpiNN$ data are analysed, especially those obtained more recently by the CLAS Collaboration at JLab, the A2 Collaboration at MAMI at Mainz, and the PION@MAX-lab Collaboration at Lund. A procedure, which accounts for the \fsi s (FSI), is applied to extract $\gnpiN$ \diff\ \crss\ from the deuteron data. The role of FSI is discussed. We also comment on the results of other works in connection with some discrepancies seen in comparing the model with experiment. The electromagnetic properties of baryon $N^\ast$ resonances, improved using the extracted $\gnpiN$ \diff\ \crss\, are presented. A model description of the cross section data from charged-pion \photo\ reactions $\gd\to\pi^{\pm}NN$ near threshold is also given. The $\gdpipp$ data are used to extract the $E_{0+}$ multipole and total \crs\ of the reaction $\gnpip$ near threshold.
\end{abstract}

\vspace{5mm}
\section{Introduction}
\label{Sec:intro}
\vspace{2mm}

The reactions of meson photoproduction on the nucleons are one of the main sources of information about the electromagnetic properties of the nucleon resonances. 
The data on both proton and neutron targets are needed to extract all the isospin amplitudes and to disentangle the isoscalar and isovector electromagnetic couplings of the various $N^\ast$ and $\Delta^\ast$ resonances. Since the neutron targets do not exist, it remains to use nuclear ones. In this case, when extracting information on the elementary reaction on the neutron from nuclear data, one should take into account the nuclear-medium effects, i.e., the \fsi\ (FSI) and Fermi-motion effects. Reactions on a deuteron target (the simplest nucleus) are the most convenient for this purpose.

A theory of pion photoproduction was constructed in the 1950's. Kroll and Ruderman~\cite{Kroll:1953vq} were the first to derive model-independent predictions in the threshold region, a so-called low energy theorem (LET), by applying gauge and Lorentz invariance to the reaction $\gNpiN$. The general formalism for this process was developed by Chew \textit{et al.}~\cite{Chew:1957tf} (CGLN amplitudes). Vainshtein and Zakharov extended the LET by including the hypothesis of a partially conserved axial current (PCAC)~\cite{Vainshtein:1972ih}. The derivation of the theorem is based on the use of the PCAC hypothesis and on the expansion of the amplitudes in powers of $k/m_{int}$ and $q/m_{int}$, where $k$ and $q$ are the pion and photon momenta and $m_{int}$ is an internal mass. This work succeeded in describing the threshold amplitude as a power series in the ratio $\kappa=m_{\pi}/m_N$ up to terms of order $\kappa^2$ ($m_{\pi}$ and $m_N$ are the averaged pion and nucleon masses, respectively). Somewhat later, Berends \textit{et al.}~\cite{Berends:1967vi} analysed the existing data in terms of a multipole decomposition and extracted the various multipole amplitudes contributing in a region up to an excitation energy of 500~MeV. These amplitudes are vital inputs to low-energy descriptions of hadron physics based on the chiral perturbation theory (ChPT)~\cite{Hilt:2013fda}. Predictions for low-energy multipoles, using a relativistic formulation of ChPT, were presented in Ref.~\cite{Bernard:1992nc}; results based on the heavy-baryon approach may be found in Ref.~\cite{Bernard:1996ti}.

Incoherent pion \photo\ on the deuteron is interesting in various aspects of nuclear physics, and particularly, provides information on the reactions on the neutron. FSI plays an important role in the analysis of the $\gNpiN$ interaction as extracted from $\gdpiNN$ data. The amplitudes for the reactions $\gamma N\to \pi N$ can be decomposed into distinct isospin $1/2$ and $3/2$ components~\cite{Bransden:1973}. As the proton-target data tend to be of a superior quality, they generally determine the isospin 3/2 components. In total, there are three isospin amplitudes ($3/2$, $p1/2$, and $n1/2$) describing the four charge channel reactions, implying that one of the four reactions is redundant and can be predicted from the other three. However, it is clear that both proton and neutron target data are required for a complete determination  and to separate the $\gamma pN^\ast$ and $\gamma nN^\ast$ photocouplings~\cite{Ireland:2019uwn}. Knowledge of the $N^\ast$ and $\Delta^\ast$ resonance photodecay amplitudes has largely been restricted to the charged states. Apart from lower-energy inverse reaction $\pi^-p\to\gamma n$ measurements, the extraction of the two-body $\gamma n\to\pi^-p$ and $\gamma n\to\pi^0 n$ observables requires the use of a model-dependent nuclear correction, which mainly comes from FSI effects.  The FSI, first considered in Refs.~\cite{Migdal:1955,Watson:1952ji}, is responsible for the near-threshold enhancement (Migdal-Watson effect) in the $NN$ mass spectrum of the meson production reaction $NN\!\to\!NNx$. In Ref.~\cite{Baru:2000hg}, the FSI amplitude was studied in detail. A series of papers, dealing with $NN$ and $\pi N$ FSI calculations for $\gdpiNN$, begins with
Refs.~\cite{Blomqvist:1977rv,Laget:1978wj,Laget:1981jq}. Laget~\cite{Laget:1978wj,Laget:1981jq}, using the $\gNpiN$ amplitude, constructed~\cite{Blomqvist:1977rv} from the Born terms and $\Delta(1232)3/2^+$ contribution,
in $\gdpiNN$ calculations with FSI terms included, succeeded in describing the available deuteron data for charged-pion \photo\ in the threshold and $\Delta(1232)3/2^+$ regions.

The topic has been developed in many papers
(see also references therein), including improvements to the $\gNpiN$ amplitude, predictions for unpolarized and polarized observables (beam, target or both) in the $\gd$ reactions. (See Refs.~\cite{Darwish:2003,Darwish:2002qu,Arenhovel:2005vh,Fix:2005vk,Laget:2006bu,Levchuk:2006vm,Levchuk:2010nb,Schwamb:2010zz,Darwish:2011,Nakamura:2017els,Nakamura:2018fad,Nakamura:2018cst} and references therein). Observables in the $\gd$ reactions have been compared with new measurements as they became available. Different models for $\gNpiN$ amplitude were used in these papers: Mainz Unitary Isobar Model MAID~\cite{Drechsel:1998hk} (Refs.~\cite{Arenhovel:2005vh,Fix:2005vk,Levchuk:2006vm}), SAID~\cite{Arndt:2002xv,Dugger:2007bt} (Refs.~\cite{Levchuk:2006vm,Levchuk:2010nb}), and MAID2007~\cite{Drechsel:2007if} (Ref.~\cite{Levchuk:2010nb}). The main uncertainties of $\gd$ calculations, as discussed in Refs.~\cite{Levchuk:2006vm,Levchuk:2010nb}, stem from the model dependence of the $\gNpiN$ amplitude. In the latest SAID~\cite{Arndt:2002xv,Dugger:2007bt} and MAID2007~\cite{Drechsel:2007if} versions, the models for $\gNpiN$ amplitudes are developed for the photon energies $\ega<2.7$~GeV~\cite{Arndt:2002xv,Dugger:2007bt} and $\ega~<~2$~GeV~\cite{Drechsel:2007if}, respectively. An essential result of the existing $\gd$ calculations shows that FSI effects significantly reduce the \diff\ \crs\ for $\pi^0 pn$ channel, mainly due to the $pn$ rescattering, and contribute much less in the charged-pion case, i.e., in $\pi^+nn$ and $\pi^-pp$ channels.

Among the above-mentioned papers, the procedure for extracting the $\gnpip$ and $\gnpin$ observables from $\gdpipp$~\cite{Aachen-Bonn-Hamburg-Heidelberg-Muenchen:1973kaj,CLAS:2017kua,CLAS:2017dco,Mattione:2011} and $\gdpipn$~\cite{Krusche:1999tv,Siodlaczek:2001mh,A2:2014fca,Dieterle:2017myg} data was specially considered in Refs.~\cite{Nakamura:2017els,Nakamura:2018cst,Nakamura:2018fad} in the framework of the model with the impulse-approximation amplitude added by FSI terms. At the end of Section~\ref{Sec:Analysis}, we shall comment some differences between their and our results discussed in Refs.~\cite{Nakamura:2018cst,Nakamura:2018fad}. 

In present paper, we review our results of theoretical study of the incoherent reactions $\gdpiNN$, performed earlier in our papers~\cite{CLAS:2017dco,Tarasov:2011ec,Chen:2012yv,Briscoe:2012ni,Tarasov:2015sta,Strandberg:2018djk,Strandberg:2017,Briscoe:2020qat}. We briefly discuss the theoretical model used for the reactions of interest. The model predictions are compared with the existing data, especially with those obtained during the last years by the CLAS Collaboration at JLab, A2 Collaboration at MAMI, Mainz, and PION@MAX-lab Collaboration at Lund. The procedure of extracting the \diff\ \crss\ of the reaction $\gnpiN$ is also considered.

\vspace{5mm}
\section{The Model}
\label{Sec:Model}
\vspace{2mm}

\begin{figure}\begin{center}
\includegraphics[width=4.4cm, keepaspectratio]{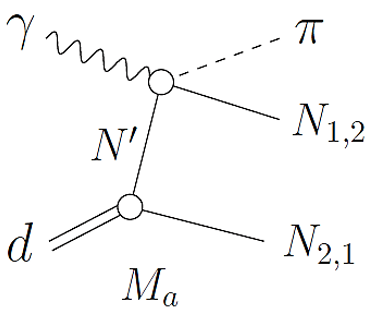}~~~~~
\includegraphics[width=4.4cm, keepaspectratio]{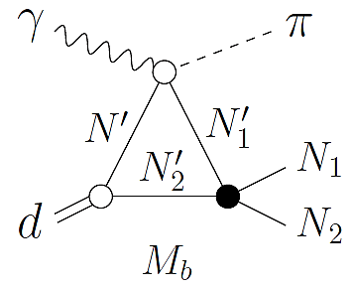}~~~~~
\includegraphics[width=4.4cm, keepaspectratio]{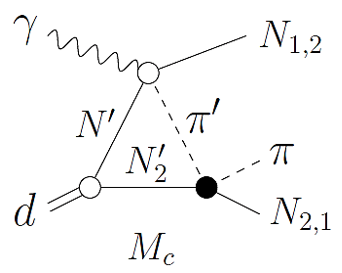}
\end{center}
\vspace{-5mm}
\caption{Feynman diagrams for the leading components of the $\gdpiNN$ amplitude. 
    (a) Impulse approximation, 
    (b) $NN$-FSI, and 
    (c) $\pi N$-FSI. 
    Filled black circles show FSI vertices. Wavy, dashed, solid, and double lines correspond to the photons, pions, nucleons, and deuterons, respectively.} \label{diag}
\end{figure}

In our model approach, the $\gdpiNN$ amplitude $M$ contains terms represented by the diagrams in Fig.~\ref{diag}, i.e.,
\be
        M_{\gd} = M_a + M_b + M_c,~~~ M_{a,c} = M^{(1)}_{a,c} + M^{(2)}_{a,c}.
\label{21}\ee
Here: $M_a$, $M_b$, and $M_c$ are the impulse approximation (IA), $NN$-FSI and $\pi N$-FSI terms, respectively. IA and $\pi N$-FSI ($M_a$ and $M_c$) diagrams include also the cross-terms between outgoing protons. Concerning the main ingredients of the model, we use the $\gNpiN$ amplitudes (upper open circles in Fig.~\ref{diag}), expressed through four spin-independent Chew-Goldberger-Low-Nambu (CGLN) amplitudes~\cite{Chew:1957tf} $F_{1-4}$, which were generated by the SAID code, using the George Washington University (GW) pion photoproduction multipoles~\cite{Arndt:2002xv,Dugger:2007bt}. For the $N\!N$-FSI and $\pi N$-FSI, we utilize the GW $N\!N$~\cite{Arndt:2007qn} and GW $\pi\!N$ (elastic$+$charge exchange)~\cite{Arndt:2006bf} amplitudes, respectively (filled circles in Fig.~\ref{diag}). For the deuteron description, we use the wave function of the Bonn potential (full model)~\cite{Machleidt:1987hj} with $S$- and $D$-wave components included. More details of the model with references are given below in the relevant parts of the text.

\vspace{5mm}
\section{The Reaction $\bfpipp$}
\label{Sec:Reaction}
\vspace{2mm}

The CLAS Collaboration data on this reaction, reported in 2009~\cite{Chen:2009sda,Chen:2010}, covers a wide energy region (1~GeV$\,<\ega<\,$3.5~GeV, laboratory frame) for the initial photon with an aim to study the reaction $\gnpip$ on the neutron. This was a motivation for our theoretical analysis of the reaction $\gdpipp$ and our formulated procedure to extract the $\gnpip$ data. We apply the model, described in Sec.~\ref{Sec:Model}, where $\pi=\pi^-$ and the nucleons $N_{1,2}$ are the protons $p_{1,2}$ in Fig.~\ref{diag}. The invariant IA amplitude is 
\be
    M^{}_a\!=\!2\sqrt{m}\,\sum_{m'}\,
    \bigl[\langle m_1|\hmgn^{(1)}|m'\rangle
    \langle m'\!,m_2 |\hPsi(\bfp_2)|m_d\rangle
    -\langle m_2|\hmgn^{(2)}|m'\rangle
    \langle m'\!,m_1|\hPsi(\bfp_1)|m_d\rangle\bigr].
\label{ma1}\ee
Here: $\mgn^{(i)}=\langle m_i|\hmgn^{(i)}|m'\rangle$ are the $\gnpip_{i}$ amplitudes; $m_{1,2}\,$($m'$) and $m_d$ are the spin states of final (virtual) nucleons and deuteron, respectively (photon polarization vector $\bfe$ is hidden in the operators $\hmgn$); $\langle m'\!,m_i|\hPsi(\bfp_i)|m_d\rangle\equiv \varphi'^+_1\hPsi(\bfp_i)\sigma_2\varphi^\ast_i$ is the The deuteron wave function (DWF) ($\varphi'$ and $\varphi_i$ are the nucleon spinors at the deuteron vertex, $\varphi^+\varphi\!=1$), and
\be
    \hPsi(\bfp)=u(p)\hat S_u\!+w(p)\hat S_w,~~~
    \hat S_u\!=\!\frac{(\bfsig\cdot\bfeps)}{\sqrt{2}},~~~
    \hat S_w\!=\half [(\bfsig\cdot\bfeps)-\frac{3}{p^2}(\bfp\cdot\bfeps)(\bfsig\cdot\bfp)]~~
    (p=|\bfp|), 
\label{de1}\ee
where $\bfeps$ is the deuteron polarization vector; $\bfsig=\{\sigma_i\}$ are the Pauli matrices; $u(p)$ and $w(p)$ are $S$- and $D$-wave parts of DWF, respectively, with normalization $\int\!\!d\bfp\,\,[u^2(p)+w^2(p)]=(2\pi)^3$.
We use the CGLN amplitudes $F_{1-4}$, defined in the c.m. frame of the process $\gNpiN$ to derive the invariant amplitudes $\mgn^{(i)}$.
Then, they are transformed to the forms $\mgn^{(i)}=\langle m_i|\hmgn^{(i)}|m'\rangle  =\langle m_i|L + i\bfK\cdot\bfsig|m'\rangle$ in the laboratory system (deuteron rest frame) to be used in Eq.~(\ref{ma1}).

The $N\!N$-FSI term $M_b$ can be written as
\be\ba{c}\dist
    M_b=\int\!\!\frac{d\bfp'_N}{(2\pi)^3}\,
    \frac{\langle\cdots\rangle}{4E'\,W\,(E'-E-i0)},
\\ \rttt\dist
    \langle\cdots\rangle=2\sqrt{m}\!\!\sum_{m',m'_1,m'_2}\!\!
    \langle m_1\!,m_2|{\hM}_{N\!N}|m'_1\!,m'_2\rangle
    \langle m'_1|\hmgn|m'\rangle
    \langle m'\!,m'_2|\hPsi(\bfp'_2)|m_d\rangle,
\ea
\label{mb1}\ee
where $W$ is the $N\!N$-system effective mass, $E=\!W/2\!=\!\sqrt{p^{\,2}_N\!+\!m^2}$, $p^{}_N\!=\!|\bfp^{}_N|$,
$\,E'\!=\!\sqrt{p^{\,\prime\,2}_N\!+\!m^2}$, $p'_N\!=\!|\bfp'_N|$, and $\,\bfp^{}_N$($\bfp'_N$) is the relative 3-momentum in the final (intermediate) $N\!N$ state. Eq.~(\ref{mb1}) for the amplitude $M_b$ is obtained in the approximation, in which the integral over the energy of the intermediate nucleon $N'_2$ is related to the residue at the nucleon pole with positive energy.

The term $\langle m_1\!,m_2|\hM_{N\!N}|m'_1\!,m'_2\rangle$ in Eq.~(\ref{mb1}) is the invariant $N\!N$(here $pp$)-scattering amplitude. Its expression is written out in Appendix (item~5) of Ref.~\cite{Tarasov:2011ec}.

According to symbolic equality $1/(E'\!-\!E\!-\!i0)=i\pi\delta(E'\!-\!E)+\!P\,(1/(E'\!-\!E))$ we can split the amplitude $M_b$ in its on- and off-shell parts and obtain
\be\ba{c}
    M_b=M^{on}_b+M^{off}_b,
\\ \rttt\dist
    M^{on}_b=\frac{ip_N}{32\pi^2 W}\int\!\!d\Omega'\,\langle\cdots\rangle,
    ~~~
    M^{off}_b=\!\frac{1}{32\pi^2 W}\int\!\!d\Omega'\!
    \oint\frac{dp'_N\,p'^2_N}{\pi E'}\frac{\langle\cdots\rangle}{E'\!-\!E},
\ea
\label{mb2}\ee
where $\oint$ denotes the principal part of the integral, and $d\Omega'=dz'd\varphi'$ ($z'=\cos\theta'$) is the element of solid angle of relative motion of the intermediate nucleons.
We also include the off-shell correction to the ${^1}S_0$ partial amplitude of $pp$-scattering in the form
\be
    \hM_{N\!N}({^1\!}S_0)=f(p'_N,p_N)\,\hM^{on}_{N\!N}({^1\!}S_0),
    ~~~f(p'_N,p_N)=\frac{p^2_N+\beta^2}{p'^2_N+\beta^2}
\label{mb3}\ee
with $\beta = 1.2\,$fm$^{-1}$~\cite{Kolybasov:1976mm} (also used in Ref.~\cite{Levchuk:2006vm}).

The $\pi N$-FSI term $M_c$ [Fig.~\ref{diag}] of the $\gd$ amplitude can be written in the form 
\be\ba{c}\dist
    M_c=M^{(1)}_c+M^{(2)}_c,~~~~
    M^{(1)}_c=\!\int\!\frac{d\bfk'_2}{(2\pi)^3}\,
    \frac{\langle\cdots\rangle}{4E'W_2\,(E'\!-E\!+i0)},
\\ \rttt\dist
    \langle\cdots\rangle=2\sqrt{m}\sum_{m',\,m'_2}
    \bigl[\,
    \langle m_1|\,\hM^{(1)}_{\gnpip}|m'\rangle
    \,\langle m_2|\,\hM^{(2)}_{\pi^- p}|\,m'_2\rangle- 
    \\ \rptt\dist
    -\langle m_1|\,\hM^{(1)}_{\gppip}|m'\rangle
    \,\langle m_2|\,\hM^{(2)}_{cex}|\,m'_2\rangle\bigr]\,
    \langle m'\!,m'_2|\,\hPsi(\bfp'_2)|\,m_d\rangle.
\ea
\label{mc1}\ee
Here, the integral over the energy of the intermediate nucleon $N'_2$ is related to the residue at the pole as in Eq.~(\ref{mb1}); $E=\!\sqrt{k^{\,2}_2\!+m^2}$,~ $E'\!=\!\sqrt{k'^{\,2}_2\!+m^2}$,~$k_2=|\bfk_2|$, and $k'_2=|\bfk'_2|$, where $\bfk'_2$ ($\bfk_2$) is the relative 3-momentum in the intermediate $\pi N$ (final $\pi N_2$) system; $\,W_2$ is the effective mass of the $\pi N_2$ system; $\hM^{(2)}_{\pi^- p}$ and $\hM^{(2)}_{cex}$ are the elastic and charge-exchange ($\pi^0 n\!\to\!\pi^- p\,$ here) $\pi N$ amplitudes, respectively; the notations $m'$, $m^{}_{1,2}$, and $m_d$ are given above. The relative sign ``-'' between two terms of $\langle\cdots\rangle$ in Eq.~(\ref{mc1}) arises from isospin antisymmetry of the DWF with respect to the nucleons. The 2nd term $M^{(2)}_c=-M^{(1)}_c$ (with permutation of the final nucleons).
Splitting the amplitude $M^{(1)}_c$ in its on- and off-shell parts, we obtain
\be\ba{c}
    M^{(1)}_c=M^{(1),on}_c+M^{(1),off}_c,
    \\ \rttt\dist
    M^{(1),on}_c=\frac{ik_2}{32\pi^2 W_2}\int\!\!d\Omega'\,\langle\cdots\rangle,
    ~~~
    M^{(1),off}_b=\frac{1}{32\pi^2 W_2}\int\!\!d\Omega'\!\oint
    \frac{dk'_2\,k'^2_2}{\pi E'}\frac{\langle\cdots\rangle}{E'\!-\!E},
\ea
\label{mc2}\ee
where $d\Omega'=dz'd\varphi'$ is the element of solid angle of relative motion in the intermediate $\pi N$ system.
The $\pi N$-scattering amplitude is described in Appendix (item~6) of Ref.~\cite{Tarasov:2011ec}.

The reaction \crs\ (unpolarized case) reads
\be
    \sigma(\gdpiNN)=J^{-1}\!\int\overline {|M_{\gd}|^2}\,d\tau_3,
    ~~~ J=4\ega m_d=4q_{\gd}\sqrt{s}.
\label{sigma}\ee
Here $m_d$ is the deuteron mass, $q_{\gd}$ is the initial relative momentum, and $\sqrt{s}$ is the total c.m. energy; $d\tau_3$ is the element of the final invariant $\pi NN$ phase space defined in the known way (see also Ref.~\cite{Tarasov:2011ec}, Appendix, Eq.~(A1)).
\vspace{5mm}

\begin{figure}\begin{center}
\includegraphics[width=14cm, keepaspectratio]{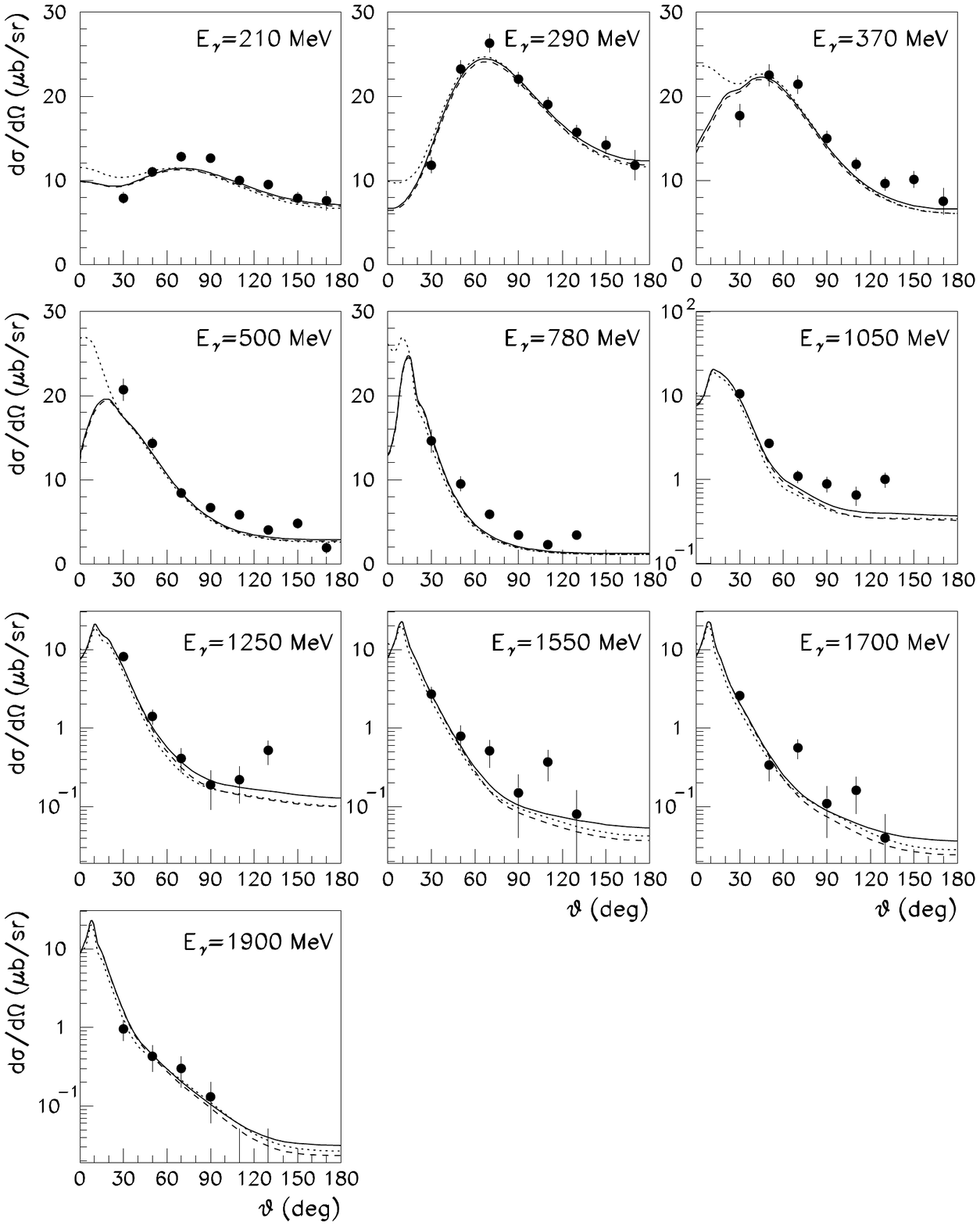}
\end{center}
\vspace{-3.2cm}

\caption{The \diff\ \crs\ $\dso$ of the reaction $\gdpipp$ in the laboratory frame at different photon energies $\ega~\le~1900$~MeV; $\theta$ is the polar angle of the outgoing $\pi^-$. Dotted curves show the contributions from the IA amplitude $M_a$ in Fig.~\ref{diag}. Successive addition of the $N\!N$-FSI and $\pi N$-FSI amplitudes $M_b$ and $M_c$ leads to dashed and solid curves, respectively. The filled circles are the data from DESY collected by the Aachen-Bonn-Hamburg-Heidelberg-M\"unchen Collaboration~\cite{Aachen-Bonn-Hamburg-Heidelberg-Muenchen:1973kaj}.}
 \label{Benz}
\end{figure}

Here we present a comparison of our model predictions with the experimental data from DESY by the Aachen-Bonn-Hamburg-Heidelberg-M\"unchen Collaboration~\cite{Aachen-Bonn-Hamburg-Heidelberg-Muenchen:1973kaj} on the \diff\ \crss\ $\dso$ versus $\theta$, where $\Omega$ and $\theta$ are solid and polar angles of outgoing $\pi^-$ in the laboratory frame, respectively, with $z$ axis along the photon beam. The results are given in Fig.~\ref{Benz}. The dotted curves are the results obtained with the IA amplitude $M_a$. Successive addition of the $N\!N$-FSI and $\pi N$-FSI amplitudes $M_b$ and $M_c$ leads to dashed and solid curves, respectively. Fig.~\ref{Benz} demonstrates a sizable FSI effect at small angles $\theta\lesssim 30^{\,\circ}$ which mainly comes from $NN$-FSI (the difference between dotted and dashed curves). Comparison of dashed and solid curves shows that $\pi N$-FSI affects the results very slightly. Note that at the energies $\ega~=~300 - 500$~MeV the effective masses of the final $\pi p$ states predominantly lie in the $\Delta(1232)$ region. Thus, the plots at $\ega~=~370$~MeV and 500~MeV of Fig.~\ref{Benz} show that the role of $\pi N$-FSI even in the $\Delta(1232)$ region is very small.
Fig.~\ref{Benz} demonstrates a reasonable description of the data~\cite{Aachen-Bonn-Hamburg-Heidelberg-Muenchen:1973kaj} on $\dso$. These data are also confirmed by the results, obtained later from the Gerasimov-Drell-Hearn experiments in Mainz~\cite{Ahrens:2010zz}. Note that the data are absent at small angles $\theta\lesssim 30^\circ$, where the FSI effects are sizeable. This is also the region of the most pronounced disagreements between the theoretical predictions of different authors~\cite{Ahrens:2010zz}.

In more detail, the role of FSI is shown in Fig.~3 of Ref.~\cite{Tarasov:2011ec} at $\ega~=~500$~MeV. There, the dashed curve is the contribution of the IA- and $NN$-FSI terms ($M_a$ and $M_b$) with only $s$-wave $N\!N$-scattering amplitude
taken into account. The dotted and solid curves have the same meaning as in Fig.~\ref{Benz}.  Thus, at small angles the $S$-wave part of NN-FSI dominates in the FSI contribution.

\vspace{3mm}
\subsection{Extraction of the $\gnpip$ \crss\ from the $\gamma d$ data}
\vspace{2mm}

The data on the deuteron target doesn't provide direct information on the \diff\ \crs\ $\dso(\gnpip)$, since the $\gdpipp$ amplitude (\ref{21}) squared $\overline{|\mgd|^2}$ can not be expressed through the $\gnpip$ amplitude squared $\overline{|\mgn|^2}$. If we neglect the FSI diagrams $M_b$ and $M_c$ [Fig.~\ref{diag}] and let the final proton
with momentum $\bfp_1$($\bfp_2$) be fast (slow) in the laboratory system and denoted by $p_1$($p_2$), then the IA diagram $M^{(1)}_a$ with slow proton $p_1$ emerging from the deuteron vertex dominates, $M^{(2)}_a$ is suppressed, and $\mgd\approx M^{(1)}_a$.  This approximation corresponds to the ``quasi-free'' (QF) process on the neutron. In this case, one can relate the \diff\ \crs\ $d\sgn/d\Omega_1$ ($\Omega_1$ is the solid angle of relative motion in the $\pi^-p_1$ pair) on neutron with that on the deuteron target as
\be
    \frac{d\sgd^{QF}}{d\bfp_2\,d\Omega_1} =
    n(\bfp_2)\,\frac{d\sgn}{d\Omega_1},~~~
    n(\bfp_2) = \frac{\ega^{\,\prime}}{\ega}\,\rho(p_2),~~~
    \frac{\ega^{\,\prime}}{\ega} = 1 + \beta\cos\theta_2,~~~
    \beta = \frac{p_2}{E_2}
\label{ne2}\ee
(see also Refs.~\cite{Blomqvist:1977rv,Laget:1978wj,Laget:1981jq}). Here: $\ega^{\,\prime}$ is the photon energy in the rest frame of the virtual neutron with momentum $p^{\,\prime}$ in the diagram $M_{a1}$ [Fig.~\ref{diag}]; the factor $\ega^{\,\prime}/\ega$ is the ratio of photon fluxes in $\gamma d$ and $\gamma n$ reactions; $\theta_2$ is the laboratory polar angle of final slow proton $p_2$ with three-momentum $\bfp_2$ ($p_2\!=\!|\bfp_2|$); $\rho(p)$ is the momentum distribution in deuteron and $\int\!\rho(p)\,d\bfp=1$. Hereafter, we use the notation $d\sgd^i/d\bfp_2d\Omega_1$, where index ``$i$'' specifies the $\gdpipp$ amplitude $\mgd^{\,i}$ used in calculations, namely $\,\mgd^{QF}\!=M^{(1)}_a$, $\mgd^{IA}\!=M_a$ or the full amplitude (without index) $\mgd=M_a+M_b+M_c$. Let us rewrite Eq.~(\ref{ne2}) in the form
\be
    \frac{d\sgd}{d\bfp_2 d\Omega_1} = n(\bfp_2)\,r\,\frac{d\sgn}{d\Omega_1},~~~
    r = r^{}_{\!P}\,r^{}_{F\!SI},~~~ r^{}_{\!P} = \frac{({\rm IA})}{({\rm QF})},~~~
    r^{}_{F\!SI} = \frac{({\rm full})}{({\rm IA})},
\label{ne3}\ee
where we use short notations (full)$\,=d\sgd/d\bfp_2d\Omega_1$ and $(i)=d\sgd^i/d\bfp_2d\Omega_1$ for $i=\,$QF and IA. Eqs.~(\ref{ne3}) enable one to extract the \diff\ \crs\ $d\sgn/d\Omega_1$ on neutron from $d\sgd/d\bfp_2d\Omega_1$, making use of the factors $n(\bfp_2)$ and $r$. Here: the factor $n(\bfp_2)$, defined in Eq.~(\ref{ne2}), takes into account the distribution function $\rho(p_2)$ and Fermi-motion in the deuteron; $r=r^{}_{\!P}\,r^{}_{F\!SI}$ is the correction coefficient, written as the product of two factors of different nature. The factor $r^{}_P$ takes into account the difference of IA and QF approximations, 
while $r^{}_{FSI}$ in Eq.~(\ref{ne3}) is the correction for ``pure'' FSI effect.

Generally for a given photon energy $\ega$, the \crs\ $d\sgd/d\bfp_2d\Omega_1$~(\ref{ne3}) with unpolarized particles and the factor $r$ depend on $p_2$, $\theta_2$, $\theta_1$, and $\varphi_1$, where $\theta_1$ and $\varphi_1$ are the polar and azimuthal angles of relative motion in the final $\pi^-p_1$ pair. To simplify the analysis, let us integrate the \diff\ \crs\ on deuteron over $\bfp_2$ in a small region $p_2<p_{max}$ and average over $\varphi_1$. Then, we define
\be
   \frac{d\sgd^i}{d\Omega_1}(\ega,\theta_1) = \frac{1}{2\pi}
   \int\!\frac{d\sgd^i}{d\bfp_2\,d\Omega_1}\,\,d\bfp_2 d\varphi_1
\label{ne4}\ee
(index ``$i$'' was introduced above). The \crs~(\ref{ne4}) depends on $\ega$ and $\theta_1$. Now calculate the same integral from the rhs of Eq.~(\ref{ne2}). Then, taking the \crs\ $d\sgn/d\Omega_1$ out of the integral $\int\!\!d\bfp_2$,
assuming $n(\bfp_2)$ to be a sharp function, we obtain
\be
    \frac{d\sgd^{QF}}{d\Omega_1}(\ega,\theta_1) =
    c\,\frac{d\asgn}{d\Omega_1},~~~ c = \int\!n(\bfp_2)\,d\bfp_2
    = 4\pi\!\!\int\limits^{p_{max}}_0\!\!\!\rho(p)p^2 dp,
\label{ne5}\ee
where $d\asgn/d\Omega_1$ is averaged over the energy $\ega^{\,\prime}$ in some region $\ega^{\,\prime}\sim\ega$. The value $c=c(p_{max})$ is the ``effective number'' of neutrons with momenta $p<p_{max}$ in the deuteron, and $c\to 1$ at $p_{max}\!\to\!\infty$. Further, we rewrite Eq.~(\ref{ne5}) in the form
\be
    \frac{d\sgd}{d\Omega_1}(\ega,\theta_1) = c\,R\,\frac{d\asgn}{d\Omega_1},
    ~~~ R = R^{}_P\,R^{}_{F\!SI},~~ R^{}_P = \frac{({\rm IA})}{({\rm QF})},~~
    R^{}_{F\!SI} = \frac{({\rm full})}{({\rm IA})},
\label{ne7}\ee
where $(i) = d\sgd^i/d\Omega_1$ ($i = \,$QF, IA) and (full)$ = d\sgd/d\Omega_1$ (the definitions are different from those in Eqs.~(\ref{ne3})). The factors $R$, $R^{}_P$, and $R^{}_{F\!SI}$ are similar to $r$, $r^{}_{\!P}$, and $r^{}_{F\!SI}$, but defined as the ratios of the ``averaged'' \crss\ $d\sgd^i/d\Omega_1$.

Finally, we replace $d\sgd/d\Omega_1$ in Eq.~(\ref{ne7}) by the $\gdpipp$ data and obtain
\be
  \frac{d\asgn^{\,exp}}{d\Omega_1}(\bar\ega,\theta_1) = c^{-1}\!(p_{max})
  \,R^{-1}\!(\ega,\theta_1)\,
  \frac{d\sgd^{exp}}{d\Omega_1}(\ega,\theta_1),
\label{ne8}\ee
where $d\asgn^{\,exp}/d\Omega_1$ is the neutron \crs, extracted from the deuteron data $d\sgd^{\,exp}/d\Omega_1$. Since $R$ is the ratio of the calculated \crss, we assume that (full)$\,\equiv d\sgd^{theor}/d\Omega_1=d\sgd^{exp}/d\Omega_1$.
The factor $R(\ega,\theta_1)$ in Eq.~(\ref{ne8}) also depends on the kinematical cuts applied. The value $\bar\ega$ in Eq.~(\ref{ne8}) is some ``effective'' value of the energy $\ega^{\,\prime} = \ega(1+\beta\cos\theta_2)$ in the range $\ega(1\pm\beta)$. At small momentum $p_2$ we have $\beta\ll 1$ and $\bar\ega\approx\ega$. This approximation also improves, since $\rho(p_2)$ peaks at $p_2=0$, where $\ega^{\,\prime} = \ega$.

Equation~(\ref{ne8}) is implied to be self-consistent, i.e., the $\gnpip$ amplitude, extracted from the $d\asgn^{exp}/d\Omega_1$, is the same as that used in calculations of the correction factor $R$. The extraction procedure
is the following. We use some ``good'' $\gamma n$ amplitude $\mgn^{(0)}$ (0th approximation) to calculate the factor $R$ in Eq.~(\ref{ne8}) and extract the $\gnpip$ amplitude $\mgn^{(1)}$ (1st approximation) from the \crs\ $d\asgn^{exp}/d\Omega_1$. If the correction is small, i.e., $|R-1|\ll 1$, then $\mgn^{(1)}$ is a good approximation for the $\gnpip$ amplitude. Otherwise, next iterations are needed. Thus, the preliminary analysis of the $R$ factor is important for the extraction procedure.

\vspace{3mm}
\subsection{Numerical Results for the $R$ Factor}
\vspace{2mm}

\begin{figure}\begin{center}
\includegraphics[width=12cm, keepaspectratio]{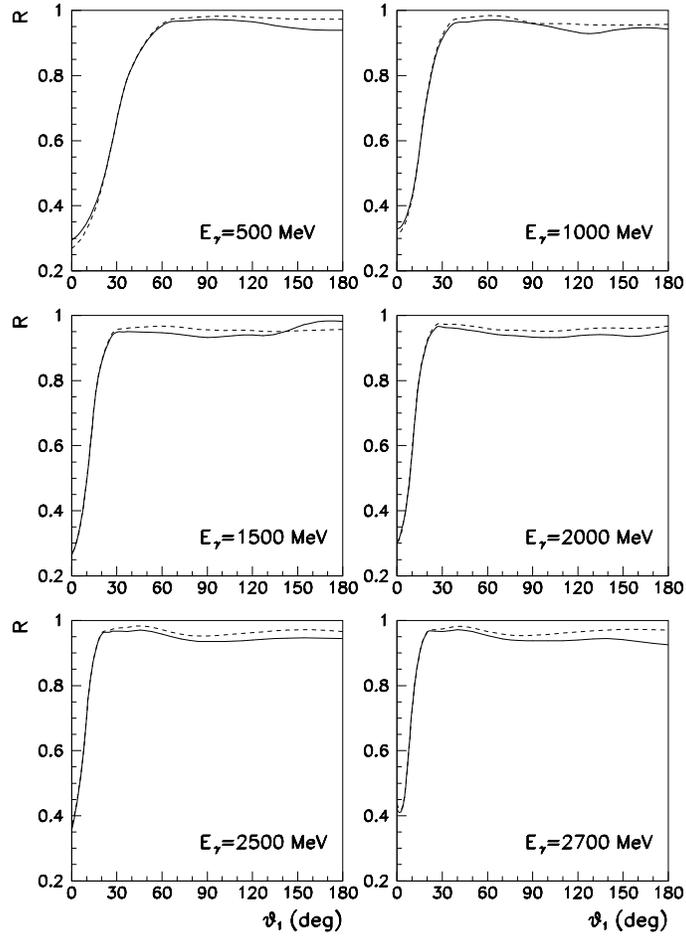}
\end{center}
\vspace{-2.3cm}

\caption{The correction factor $R$, defined by Eq.~(\ref{ne7}), where $\theta_1$ is the polar angle of the outgoing $\pi^-$ in the rest frame of the pair $\pi^-+$fast proton. The kinematical cut~(\ref{cut}) is applied. The solid (dashed) curves are obtained with both $\pi N$- and NN-FSI (only NN-FSI), taken into account.}\label{fig:R}
\end{figure}

We present in Fig.~\ref{fig:R} the model results for the correction factor $R$, defined in Eq.~(\ref{ne7}), at several photon energies $\ega$ in the range ($1000 - 2700$)~MeV. Here, we use the cuts, similar to those applied
to the CLAS data events~\cite{Chen:2010}, and select configurations with
\be
  |\bfp_2|<200~MeV/c<|\bfp_1|,
\label{cut}\ee
where $\bfp_1(\bfp_2)$ is the three-momentum of fast (slow) final proton in the laboratory system.
The solid curves show the results for $R$, where the \diff\ \crs\ (full) in Eq.~(\ref{ne5}) takes into account the full amplitude $M_a\!+M_b\!+M_c$ in Eq.~(\ref{21}). The dashed curves were calculated, excluding the $\pi N$-FSI term $M_c$ from the (full) \crs. The main features of the results in Fig.~\ref{fig:R} are
\begin{enumerate}
\item A sizeable effect is observed in some region close to $\theta_1\!=0$, which narrows as the energy $\ega$ increases;
\item The correction factor $R$ is close to 1 (small effect) in the wide angular region.
\end{enumerate}
%
Since $R$ consist of two factors $R^{}_{\!P}$ and $R^{}_{F\!SI}$, their values also are interesting. Both factors are presented in Ref.~\cite{Tarasov:2011ec} [Figs.~5(a) and 5(c)] for $\ega~=~1000$~MeV and 2000~MeV, and $R^{}_P\ne 1$ at small angles. This can be naturally understood. Since $R^{}_P$ is the correction for the 2nd (``suppressed'') IA amplitude
$M^{(2)}_a$, one expects $M^{(1)}_a\sim M^{(2)}_a$ and $R^{}_P\ne 1$ at $\bfp_1\sim\bfp_2$.  The probability of such configuration increases at $\theta_1\!\to\!0$.

The dominant role of the $S$-wave NN rescattering in the FSI effect was marked above. This contribution to the factor $R$ also is presented in Ref.~\cite{Tarasov:2011ec} [Figs.~5(b) and 5(d) therein]. There, the solid curves mean the same as in Fig.~\ref{fig:R}, i.e., the total results; the dashed curves show the values $R$, where $R^{}_{FSI}$ takes into account only the the correction from the $S$-wave part of NN-FSI. Comparison of the curves show that the FSI effect mostly comes from the $S$-wave part of $pp$-FSI. The $S$-wave NN-FSI effect is important in some region $\bfp_1\sim\bfp_2$, i.e., at small angles. Obviously, the results for $R$ are sensitive to the kinematical cuts.


\vspace{3mm}
\subsection{Factor $R$ and Glauber Approximation}
\vspace{2mm}

At large angles $\theta_1$, where FSI effects are small ($R\sim 1$), we have the rescattering of fast pion and nucleon on the slow nucleon-spectator with small momentum transfer. Then, we may estimate the FSI amplitudes in the Glauber approach~\cite{Frankfurt:1994kt}, if the laboratory momentum of rescattering particle $\gg \bar p$ (typical value in deuteron). For $N\!N$-FSI, this condition gives $\sin\theta_1\gg{\bar p\,}W_1/m\ega$, where $W_1$ is the $\pi^-p_1$ effective mass. Taking $\bar p = 150$~MeV$/c$, we get $\theta_1\!\gg 15.4^{\,\circ}(10^{\,\circ})$ for $\ega\! =~ 1000~(2000)$~MeV.
The high-energy $hN$-scattering amplitude ($h=\pi,N$) can be written as
    $M^{}_{h\!N}=2ip\,W\sigma^t_{h\!N}\exp(bt)$, 
%
where $p$, $W$, $t$, $b$, and $\sigma^t_{hN}$ are the relative momentum, $h\!N$ effective mass, the four-momentum transfer squared, slope, and total $h\!N$ \crs, respectively. The amplitude is assumed to be pure imaginary, and spin-flip term is neglected. We retain only the $S$-wave part of DWF, $u(p)$, and neglect the ``suppressed'' term $M^{(2)}_a$ of the IA amplitude $M^{}_a$ in Eq.~(\ref{21}). Finally, the FSI correction factor is $R=|M_a\!+\!M_b\!+\!M_c|^2/|M_a|^2$, and we obtain (see details in Ref.~\cite{Tarasov:2011ec})
\be
    R=R^{}_{FSI}=\left(\frac{u(0)-0.25\,
    (\sigma^t_{N\!N}\!+\!\sigma^t_{\pi N})\,J}{u(0)}\right)^2\!\approx 0.95,
    ~~~~J=\!\int\!\frac{d^2p_{\perp}}{(2\pi)^2}\,u(p_{\perp}\!)\,e^{bt}.
\label{gla}\ee
A number of approximations were made here. The factor $\exp(bt)$ is smooth in comparison with a sharp DWF $u(p_{\perp})$ in the integral $J$~(\ref{gla}), and we used $b\!=\!0$ in calculations. Considering the proton-spectator to be very slow, we evaluated the IA term $M_a$ at $u(p_2)\sim u(0)$. A typical values $\sigma^t_{N\!N}\approx 45$~mb and $\sigma^t_{\pi N}\approx 35$~mb at laboratory momenta $p_{lab}\sim 1 - 1.5$~GeV$/c$ and the $S$-wave part of Bonn (full model) DWF~\cite{Machleidt:1987hj} were used in Eqs.~(\ref{gla}).


Our simplified Glauber-type calculations give only qualitative estimation. However, the value of the FSI correction factor $R$ given in Eq.~(\ref{gla}) is in a reasonable agreement with that obtained with our full dynamical model (solid curves in Fig.~\ref{fig:R} at large angles).
The analysis~\cite{JeffersonLabHallA:2002vjd} of the reaction $\gdpipp$ at high energies, based on the approach of Ref.~\cite{Gao:1996mg}, gave the Glauber FSI correction of the order of 20\%. Similar values 15\% -- 30\% for the effect in the same approach were obtained in Refs.~\cite{Chen:2009sda,Chen:2010}, while our estimation~(\ref{gla}) gives smaller value $\sim 5\%$. Let us point out the difference of the approaches used. We use the diagrammatic technique. The approach used in Refs.~\cite{Chen:2009sda,Chen:2010,JeffersonLabHallA:2002vjd,Gao:1996mg} considers a semi-classical propagation of final particles in the nuclear matter. Its applicability to the deuteron case is rather questionable.

\vspace{3mm}
\subsection{Analysis of the $\gdpipp$ data}
\vspace{2mm}

In Ref.~\cite{Chen:2012yv}, the procedure discussed above is applied to extract the $\gnpip$ \diff\ \crs\ at $\ega~=~1.0 - 2.7$~GeV from the CLAS data~\cite{Chen:2009sda,Chen:2010}. The obtained results are in agreement with previous measurements (see Refs.~[25, 26, 27] in Ref.~\cite{Chen:2012yv}) at $\ega~=~1.15 - 1.90$~GeV displayed in Fig.~3 of Ref.~\cite{Chen:2012yv}). This CLAS data extend the results to higher energies (up to $\ega~=~2700$~MeV in Fig.~4 of Ref.~\cite{Chen:2012yv}) with more complete angular coverage. The partial-wave analysis (PWA) of these data combined with previous ones approximately confirmed the results obtained before (solution SAID SN11~\cite{Workman:2011vb}) on the neutron helicity amplitudes $A_{1/2}$ and $A_{3/2}$ for the $N(1440)1/2^+$, $N(1520)3/2^-$, and $N(1675)5/2^-$ states. The results for other states essentially depend on the details of the PWA version and are not stable.

\vspace{2mm}
In Ref.~\cite{Briscoe:2012ni}, the $\gnpip$ \diff\ \crss\ have been extracted from MAMI-B measurements of $\gdpipp$ in the $\Delta$-isobar region, accounting for the $N\!N$ and $\pi N$ FSI effects. These \diff\ \crss\ , corrected
for FSI, are given in Fig.~5 of Ref.~\cite{Briscoe:2012ni} for the photon energies $\ega~=~301 - 455$~MeV and are in a good agreement with predictions of the previous multipole analyses SN11~\cite{Workman:2011vb} and
MAID2007~\cite{Drechsel:2007if}. The new data combined with previous ones (see references in Ref.~\cite{Briscoe:2012ni}) were used in the revised analysis. Changes to the multipoles tended to be small.

\vspace{2mm}
In Ref.~\cite{CLAS:2017dco,Mattione:2011}, quasifree $\gamma d\!\to\!\pi^-p(p)$ \diff\ \crss\ have been measured with CLAS at photon beam energies $\ega~=~0.445 - 2.51$~GeV (corresponding to $W = 1.311 - 2.366$~GeV) for pion center-of-mass angles $\cos\theta^{cm}_{\pi} = -0.72 - 0.92$. Statistics in this precision experiment increased by a factor of $\sim 10$ compared to previous measurements~\cite{Briscoe:2012ni,Chen:2009sda}.  Fig.~10 and 11 of Ref.~\cite{CLAS:2017dco} show the extracted $\gnpip$ \diff\ \crss\ compared against previous measurements and available PWA solutions.
This CLAS data are systematically lower than some previous results (see references in Ref.~\cite{CLAS:2017dco}) in several energy bins at $\ega~<~1400$~MeV and are in excellent agreement with previous CLAS results~\cite{Briscoe:2012ni} at $\ega~>~1$~GeV. There is also a discrepancy in the behaviour at forward angles between the CLAS data and SLAC (see Ref.~[23] in Ref.~\cite{CLAS:2017dco}) at $\ega~<~800$~MeV, since the CLAS data rises more sharply at forward angles. Note that for these energies the CLAS data at small angles $\theta^{cm}_{\pi}$ falls into the region where the FSI-correction $R$ (Fig.~14 there) rapidly decreases. This gives a sharp rising effect in the extracted $\gnpip$ \crs\ since $\dso^{cm}_{\pi}\sim R^{-1}(\ega,\theta^{cm}_{\pi})$ according to 
Eq.~(5) of Ref.~\cite{CLAS:2017dco}. As the energy increases the range of angles with rapid $R$ behaviour narrows down close to the value $\theta^{cm}_{\pi}=0$ and doesn't affect the CLAS data obtained at $\cos\theta^{cm}_{\pi}<0.92$. This indicates the need for testing and possible further improvement of the model for the reaction $\gdpipp$ at small angles $\theta^{cm}_{\pi}$ where the FSI effects are significant.

A multipole analysis of this new CLAS data on the $\gnpip$ \crss\ $\dso^{cm}_{\pi}$ was also carried out~\cite{CLAS:2017dco} and results for several multipole amplitudes were obtained (see Figs.~17-19 there). A number of photodecay amplitudes $N^\ast\to\gamma n$ were extracted at their pole positions. This was the first pole determination of the excited neutron multipoles for the $N(1440)1/2^+$, $N(1535)1/2-$, $N(1650)1/2-$, and $N(1720)3/2+$ resonances. The values of the neutron helicity amplitudes $A_{1/2}$ and $A_{3/2}$ for these $N^\ast$ states are presented in Table~I with the results of previous analysis (see Ref.~\cite{CLAS:2017dco} and references therein). These new results have improved our knowledge of the neutral resonance properties.

\vspace{3mm}
\section{Analysis of the $\gdpipn$ Data}
\label{Sec:Analysis}
\vspace{2mm}

Here we use the invariant $\gN\to\pi^0 N$ amplitude in the form
\be
    M^{}_{\gn}\! = \!8\pi W\,A_{\gn},~~
    A_{\gn}\! = \varphi^+_f\hA_{\gn}\varphi^{}_i = A_v\pm A_s,~~
    A_{v,s}\! = \varphi^+_f(L_{v,s}\! + i\bfK_{v,s}\cdot\bfsig)\varphi_i.
\label{pipn0}\ee
Here, $W\! = \!\sqrt{s}$; the upper (lower) sign ``$\pm$'' correspond to the $\gppip$ ($\gnpin$) channel; $A_v\,(A_s$) is the isovector (isoscalar) amplitude; $\varphi_i$ ($\varphi_f$) is the spinor of the initial (final) nucleon.
The IA amplitudes for the $\gdpipn$ channel can be written as
\be\ba{l}
    M^{}_{a1}\! = c\,\varphi^+_1\hA^{(1)}_{\gn}
    \hPsi(\bfp_2)\varphi^c_2,~~~~ c = 16\pi W\sqrt{m},
    \\ \rptt\dist
    M^{}_{a2}\! = c\,\varphi^+_2\hA^{(2)}_{\gn}
    \hPsi(\bfp_1)\varphi^c_1
    = -\!2\sqrt{m}\,\varphi^+_1\hPsi^c(\bfp_1)\hA^{(2)c}_{\gn}\varphi^c_2,
\ea
\label{pipn1}\ee
where $\varphi^c\equiv\sigma_{2}\varphi^\ast$, $\hat A^c\equiv\sigma_2\hat A^T\sigma_2$ ($\sigma^c_i=-\sigma^{}_i$),
$\hPsi^c(\bfp)=-\hPsi(\bfp)$. The other notations are similar to those in Sec.~\ref{Sec:Reaction}. Below, we shall consider two cases, where $N_{1,2} = p,n$ (1st case), and $N_{1,2}=n,p$ (2nd one). Thus, $\hA^{(1)}_{\gN}$ and $\hA^{(2)}_{\gN}$ are the $\gppip$ and $\gnpin$ amplitudes in the 1st case, and otherwise in the 2nd one. Here we will also apply the following simplifications. Neglect the $\pi N$-FSI diagram $M_c$, which contribution is relatively small at $\ega$ above 200~MeV~\cite{Darwish:2003,Arenhovel:2005vh,Fix:2005vk,Levchuk:2006vm}. In the $N\!N$-FSI term $M_b$ we leave only the $S$-wave $pn$-scattering amplitude (with both isospins $0$ and $1$). The elementary $\gNpiN$ amplitude is taken out of the loop integral in the diagram $M_b$ (see details in Ref.~\cite{Tarasov:2015sta}).

In our approximation, the $NN$-FSI term $M_b$ reads
\be
    M_b = c\,\varphi^+_1\Bigl[f^{(0)}_{pn}(p)
    \bigl(L_v\bfL\! - [\bfK_v\!\times\!\bfL]\bigr)\cdot\bfsig
    \pm if^{(1)}_{pn}(p)(\bfK_s\!\cdot\!\bfL)\Bigr]\varphi^c_2,
\label{pipn3}\ee
where the upper (lower) sign correspond to the case $N_{1,2} = p,n(n,p)$. Here, $f^{(0,1)}_{pn}(p)$ are the on-shell $S$-wave $pn$-scattering amplitudes with isospins $0$ and $1$, defined as
\be
    f^{(0,1)}_{pn}(p) = (-a^{-1}_{0,1} + \half r^{}_{0,1}p^2 - ip)^{-1},
\label{pipn4}\ee
and expressed through the scattering lengths $a^{}_{0,1}$ and effective radii $r^{}_{0,1}$. We use the known values~\cite{Landau:1977}: $a_0=5.4$~fm, $r_0=1.7$~fm, $a_1=-24$~fm, and $r_1=2.7$~fm. The three-vector $\bfL$ in Eq.~(\ref{pipn3}) contains the loop integral in the amplitude $M_b$ and is given by the relations
\be
\bfL\!=[J(ip)-J(-\beta)]\,\bfeps,~~~
J(a)\equiv\!\int\!\!\frac{d^3\bfr}{2\pi r^2}
\exp(ar\!+\!i\bfdeL\cdot\bfr)\frac{u(r)}{\sqrt{2}},~~~
\bfdeL=\half(\bfp_1\!+\!\bfp_2).
\label{pipn5}\ee
Here: $2\bfdeL$ is the laboratory three-momentum of the final $N\!N$ system; $u(r)$ is the $S$-wave part of the DWF in $\bfr$-picture, normalized to $\int[u^2(r)\!+w^2(r)]^{\,}dr=4\pi$; $\beta$ is the off-shell parameter, introduced in Eq.~(\ref{mb3}). We neglect the $D$-wave part $w(r)$ of the DWF in Eq.~(\ref{pipn5}) to simplify calculations because the loop integral in the term $M_b$ is dominated by the region of small momenta in the deuteron vertex, where the $D$-wave contribution is relatively small. With the DWF, parametrized in Ref.~\cite{Machleidt:1987hj} (Bonn potential, full model), the integrals in Eq.~(\ref{pipn5}) are carried out analytically (see Ref.~\cite{Tarasov:2015sta}, Appendix~A).
\begin{figure}\begin{center}
\includegraphics[width=12cm, keepaspectratio]{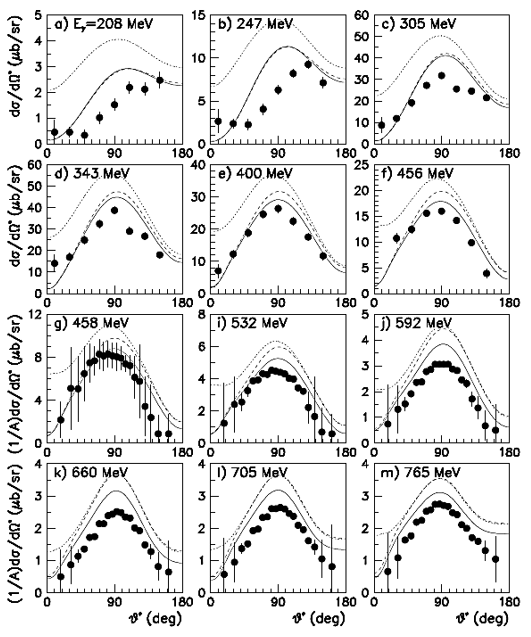}
\end{center}
\vspace{-0.7cm}

\caption{The \diff\ \crss\ of the $\gdpipn$ for several values of the photon-beam laboratory energy $\ega$ vs. $\theta^\ast$ (the angle $\,\theta^\ast$ is defined in the text). The curves show the contributions: dotted -- from the IA amplitudes $M_{a1,a2}$ (\ref{pipn1}); dashed -- from the amplitude $M_{\gd}\!=\!M_{a1}\!+\!M_{a2}\!+\!M_b$ with the $N\!N$-FSI term $M_b$ (\ref{pipn3}); solid -- from the amplitude $M_{\gd}$ with the off-shell correction~(\ref{gNoff}) included. The filled circles: in the plots ($a$)-($f$) -- the data from MAMI~\cite{Krusche:1999tv} (error bars include statistical uncertainties only); in the plots ($g$)-($m$) -- the data from MAMI as well~\cite{A2:2014fca} (error bars include statistical and systematic uncertainties in quadrature), multiplied by $1/A$ ($A=2$ for the deuteron).}
\label{pipn:crs}
\end{figure}

The results of the model for \diff\ \crss\ $\dsdo^\ast$ of the reaction $\gdpipn$ for several photon laboratory energies $\ega$ are compared in Fig.~\ref{pipn:crs} to experimental data from Refs.~\cite{Krusche:1999tv,A2:2014fca}.
Here, $\theta^\ast$ is the polar angle of the outgoing $\pi^0$ in the so-called $\gN$ c.m. frame, defined as the c.m. system of the incident photon and a nucleon at rest in the laboratory frame with z-axis directed along the photon momentum.
The dotted curves show the contributions from the IA term $M_a\! = M_{a1}\! + M_{a2}$, while the dashed ones represent the results obtained with the full amplitude $M_{\gd}\! = M_{a1}\! + M_{a2}\! + M_b$. The results show an important role of the $N\!N$-FSI, which essentially decrease the \crss\ at low energies in line with other papers~\cite{Darwish:2003,Arenhovel:2005vh,Fix:2005vk,Levchuk:2006vm}. Here, in the $\Delta(1232)$ region, the
$N\!N$-FSI effect in the $\pi^0pn$ channel is essentially more important than in the case of charged-pion production (see Fig.~\ref{Benz}). This is because of a cancellation effect in the full amplitude $M_{\gd}\! = M_{a1}\! + M_{a2}\! + M_b$ due to the orthogonality between the initial deuteron and the final $pn$ plane-wave states. In the $\Delta(1232)$ region, the reaction $\gdpipn$ proceeds predominantly through the isovector $\gNpiN$ amplitude, producing the final $pn$ system in the isoscalar state. The cancellation effect enhances when the momentum transfer from the initial photon to the final pion decreases, and turns to be maximal at zero angle of the outgoing pion. The role of the orthogonality effect was also discussed in Refs.~\cite{Arenhovel:2005vh,Fix:2005vk,Fix:2019txp}. At higher energies [plots($i$)--($l$)], the FSI effect is very small, except the region of small angles $\theta^\ast$, where it is sizeable.
Our model predictions (dotted curves) sizeable overestimate the data, but the shape of the \diff\ \crss\ in the main is reproduced. We address these discrepancies of our calculations to the model approximations applied. The results of Refs.~\cite{Darwish:2003,Arenhovel:2005vh,Fix:2005vk,Levchuk:2006vm,Levchuk:2010nb} also shows some overestimation of the data on $\gdpipn$ in the $\Delta(1232)$ energy region. The question about the source of this discrepancy remains open~\cite{Fix:2019txp}.
Some $\pi N$-FSI contribution reducing the \crss\ at $\ega~\sim~500 - 720$~MeV is seen in the results of Nakamura~\cite{Nakamura:2018fad} (see Figs.~4 and 5 therein). Their predictions also overestimate data on the $\gdpipn$ \diff\ \crss\ (see Fig.~2 in Ref.~\cite{Nakamura:2017els}). 

Let us try to improve the theoretical description, introducing the off-shell correction to the elementary $\gnpin$ amplitudes in the diagrams $M_{a1}$ and $M_{a2}$. We multiply the $\gdpipn$ amplitude by the off-shell correction factor 
\be
    F(\qga,\qga')=\frac{\Lambda^2\!+\qga^2}{\Lambda^2\!+\qga'{^{\!2}}}.
\label{gNoff}
\ee
Here: $\qga\!=(W^2\!-m^2)/2W$ and $\qga'\!=(W^2\!-p'^2)/2W$,
where $\qga$ ($\qga'$) is the initial relative momentum in the $\gnpin$ reaction with a free (virtual) initial nucleon; $W$ is the effective mass $W$ of the final $\pi N$ pair; $m$ is the nucleon mass; $p'$ is the four-momenta of the virtual nucleon in the IA diagrams $M_{a1,a2}$ in Fig.~\ref{diag}. For simplicity, we neglect this correction in the $N\!N$-FSI term, where the nucleon momenta in the deuteron vertex are effectively small in the loop integral, and the off-shell effect is expected to be small. We also expect that the IA contributions at high energies [Fig.~\ref{pipn:crs}, plots ($i$)--($l$)] are overestimated in the model and should be suppressed.
Solid curves in Fig.~\ref{pipn:crs} performs the off-shell corrected results from the full amplitude $M_{\gd}$ with $\Lambda=1$~fm$^{-1}$ in Eq.~(\ref{gNoff}). The role of the off-shell correction is negligible
at low energies [Fig.~\ref{pipn:crs}, plots ($a$)--($c$)], visibly decreases the \crss\ at higher energies, and the theoretical description looks better, but still overestimates the data. Varying the value of $\Lambda$~(\ref{gNoff}), one can not further improve the description of the data.

\vspace{1mm}
Now discuss the procedure of extracting the $\gN\!\to\pi^0 N$ \crs\ from the deuteron data. Let $N_1$ and $N_2$ are, respectively, fast and slow final nucleons in laboratory frame. Using the notations from Sec.~\ref{Sec:Analysis}, we write
\be
    \frac{d\sigma}{d\Omega}(\gN\!\to\pi^0 N) = \frac{r}{n(\bfp_2)}\,
    \frac{d\sigma^{exp}_{\gd}}{d\Omega\,d\bfp_2},
    ~~~~~
    r = \frac{d\sigma^{QF}_{\gd}}{d\Omega\,d\bfp_2}
    \bigg/\!\frac{d\sigma_{\gd}}{d\Omega\,d\bfp_2}.
\label{r1}\ee
Here, the \diff\ \crss\ $d\sigma^{QF}_{\gd}/d\Omega/d\bfp_2$ and $d\sigma_{\gd}/d\Omega/d\bfp_2$ are calculated with the amplitudes $M_{a1}$ (dominant IA term) and $M_{\gd}\! = M_{a1}\! + M_{a2}\! + M_b$, respectively; $d\Omega$ is the solid angle element of the outgoing $\pi^0$ in the $\pi^0 N_1$ c.m. frame with $z$-axis along the photon beam. The factor $r$ in Eqs.~(\ref{r1}) gives the FSI correction to the $\gppip$ ($\gnpin$) \diff\ \crs\ on the proton (neutron) in the case $1$~($2$). Integrating over spectator momentum $\bfp_2$, we obtain
\be
    r\to R_{p,n} = \frac{d\sigma^{QF}_{\gd}}{d\Omega}
    \bigg/\!\frac{d\sigma_{\gd}}{d\Omega}
\label{r2}\ee
for the correction factor $R_p$ ($R_n$) in the case $1$ ($2$). In the $\Delta(1232)$ region, where only isovector part
contributes to the $\gN\!\to\pi^0 N$ amplitude, all the terms in the $\gdpipn$ amplitude are the same in the cases $1$ and $2$. Thus, in this region we have $R_p=R_n$.
\begin{figure}\begin{center}
\includegraphics[width=14cm, keepaspectratio]{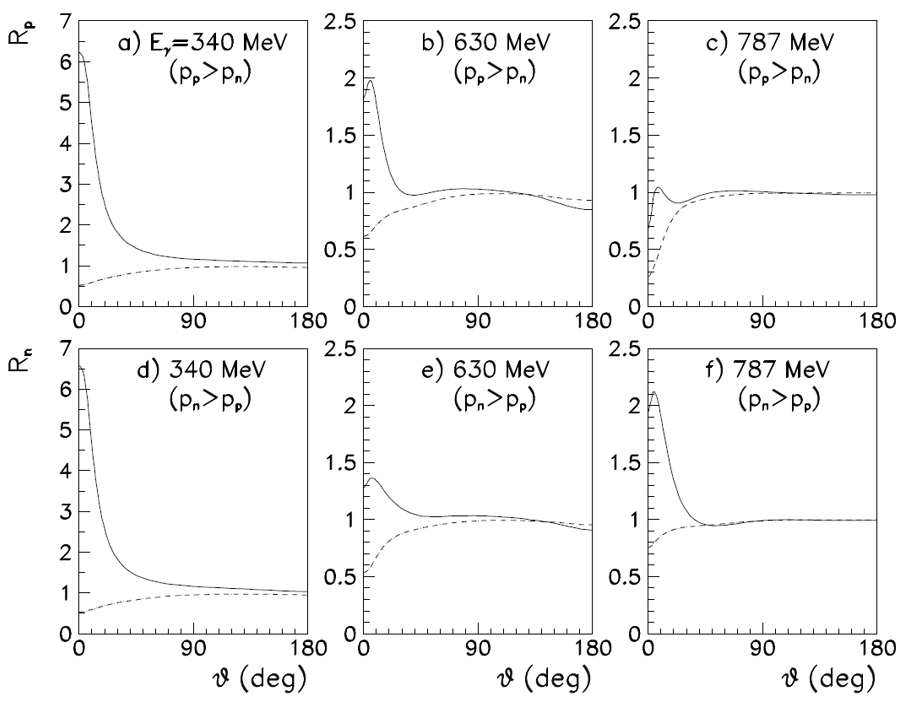}
\end{center}
\vspace{-0.5cm}

\caption{The correction factors $R_p$ [($a$)--($c$)] and $R_n$ [($d$)--($f$)], calculated according to Eq.~(\ref{r2}) from the reactions $\gdpipn$ with fast proton (slow neutron) and fast neutron (slow proton), respectively; Left, middle and right plots -- the results for $\ega~=~340$~MeV ($\Delta(1232)1/2^+$), 630~MeV ($N(1440)1/2^+$), and 787~MeV ($N(1535)1/2^-$), respectively. The numerator $d\sigma^{QF}_{\gd}\!/d\Omega$ in Eq.~(\ref{r2}) is obtained from the leading IA amplitude $M_{a1}$. Successive addition of the ``suppressed'' IA term $M_{a2}$ and $N\!N$-FSI term $M_b$, when calculating the denominator $d\sigma_{\gd}\!/d\Omega$ in Eq.~(\ref{r2}), leads to dashed and solid curves, respectively.}
\label{pipn:R}
\end{figure}

The theoretical predictions for the factors $R_p$ and $R_n$, defined in Eq.~(\ref{r2}), are given in Fig.~\ref{pipn:R} at several photon energies. Here, we define the ``fast'' and ``slow'' nucleons by the inequality $|\bfp_1|\!>\!|\bfp_2|$, and the \diff\ \crss\ in Eq.~(\ref{r2}) are integrated over this kinematic region. Hereafter, we do not apply the off-shell correction Eq.~(\ref{gNoff}) to the $\gNpiN$ amplitude. The polar angle $\theta$ of the outgoing pion is defined in the $\pi^0p$ (case~$1$) or $\pi^0n$ (case~$2$) c.m. frame. The types of curves in Fig.~\ref{pipn:R} specify the calculation of the the denominator $d\sigma_{\gd}\!/d\Omega$ in Eqs.~(\ref{21}). The dashed curves are obtained with the IA amplitude $M_a\!=M_{a1}\!+\!M_{a2}$ for this \crs, and the ``suppressed'' IA term $M_{a2}$ alone already produces a visible deviation $R_{p,n}\ne 1$, which increases to small angles. Addition of the $N\!N$-FSI term $M_b$ leads to the solid curves and considerably affects the results.
Both terms ($M_{a2}$ and $M_b$) essentially affect the results at small angles $\theta$ (pions emitted at forward angles), where the configuration with small relative momenta between the final-state nucleons dominates, and this $\theta$ region narrows with the increasing photon energy. At higher angles, both effects are negligible, and $R_{p,n}\approx 1$.
At $\ega\!=~340$~MeV [plots ($a),(d$)] we obtain a large effect at $\theta\sim 0$, where $R_{p,n}\sim 6$. This result in the $\Delta(1232)$ region arises due to the cancellation effect in the full amplitude $M_{\gd}\!=M_{a1}\!+M_{a2}\!+M_b$ mentioned above, which decreases the denominator $d\sigma_{\gd}\!/d\Omega$ in Eq.~(\ref{r2}).
Fig.~\ref{pipn:R} [plots ($a),(d$)] also shows that $R_p~=~R_n$ to a good accuracy in the $\Delta(1232)$ region. At higher energies (630 and 787~MeV), we observe $R_p\!\ne\!R_n$, since both isovector and isoscalar parts of the $\gN\!\to\pi^0 N$ amplitudes contribute. This difference is considerable at small angles $\theta$ where the role of the ``suppressed'' IA term $M_{a2}$ and $N\!N$-FSI is enhanced.
\vspace{1mm}

The model was applied to extract the $\gnpin$ \crs\ from the Mainz A2 experiment on the deuteron target~\cite{A2:2019yud}. The $\gnpin$ \diff\ \crss\ were obtained at the photon-energy range $290 - 813$~MeV
($W_{\pi^0n}\!=\!1.195 - 1.553$~GeV) and the pion c.m. polar production angles, ranging from 18$^{\circ}$ to 162$^{\,\circ}$. To suppress the main background for the neutral channel, coming from the reaction $\gppip$, the events with high-energy protons were discarded from the analysis. The $pn$-FSI correction, included in the data analysis, means taking into account each event with a weight as
\be
    R = \overline{|M_{a1}|^2}/\,\overline{|M_{\gd}|^2}
\label{r3}\ee
($M_{\gd}\! = M_{a1}\!+M_{a2}\!+M_b$), where $\overline{|M_{a1}|^2}$ and $\overline{|M_{\gd}|^2}$ are the amplitudes squared, averaged over spins and calculated for the kinematics of the event. Further event handling was carried out under the assumption that the reaction mechanism is determined by the IA diagram $M_{a1}$ in Fig.~\ref{diag}, where $N_{1,2} = n,p$.
\begin{figure}[htb]
\begin{center}
\includegraphics[width=10cm, keepaspectratio]{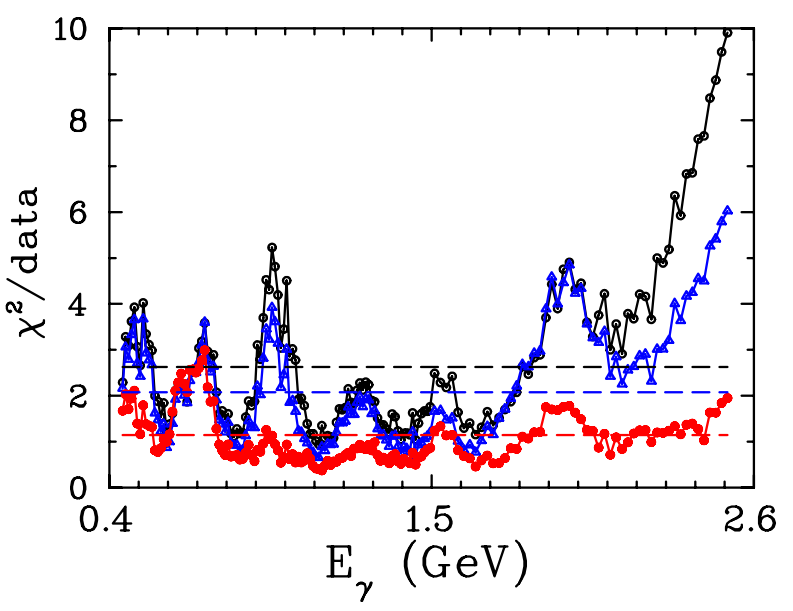}
\end{center}
\vspace{-0.5cm}

\caption{Comparison of the SAID prediction PR15~\cite{A2:2015mhs} applied to the new CLAS cross sections for the reaction $\gamma n\to\pi^-p$ with (blue filled triangles) and without FSI corrections (black open circles), and the SAID MA27~\cite{CLAS:2017dco} (red full circles) solution obtained after adding these data with FSI corrections into the fit. The solid lines connecting the points are included only to guide the eye. Shown are the fit $\chi^2$ per data point values averaged within each energy bin $E_\gamma$, where the horizontal dashed lines (blue (black) for PR15 and red for MA27).}
\label{chi2}
\end{figure}

The extracted \diff\ \crss\ on the reaction $\gnpin$ are compared in Figs.~7 of Ref.~\cite{A2:2019yud} ($\ega~=~512$, 610, 690, and 789~MeV) together with previous measurements and predictions of several multipole fits (see references therein). In general, these new results are in a reasonable agreement with other data as well as with previous A2 measurements~\cite{A2:2018jcd} except those at $\ega~=~789$~MeV, where the new \crss\ values are noticeably above those from Ref.~\cite{A2:2018jcd}.  In Figs.~9 and 10 of Ref.~\cite{A2:2019yud}, new data on the $\gnpin$ \diff\ \crss\ at several $\ega$ values (27 plots) and predictions of different multipole fits are presented (Fig.~9: $\ega~=~290 - 670$~MeV; Fig.~10: $\ega~=~690 - 813$~MeV). New results of the multipole fit labeled MA19, which includes a fit of the new data, and the previous MA27~\cite{CLAS:2017dco} solution are also shown in these Figures. Both fits, MA19 and MA27, are in a reasonable agreement with each other at  energies below $\ega~\sim~600$~Mev, but their difference becomes significant at higher energies.  Fig.~\ref{chi2} shows the improvement of the fit of new CLAS measurements including FSI corrections vs the world database~\cite{CLAS:2017dco}.

Nakamura \textit{et al.}~\cite{Nakamura:2018cst,Nakamura:2018fad} discussed the differences between their and our approaches on several points. \\
(i) First of all: they included the off-shell behavior of the $\gN,\pi N\!\to\pi N$ and $NN\!\to NN$ amplitudes involved in the $\gdpi$ terms. They use the off-shell-dependent $\gN,\pi N\!\to\pi N$ and $NN\!\to NN$ amplitudes generated from the ANL-Osaka dynamical coupled-channels model~\cite{Kamano:2013iva} and CD-Bonn potential~\cite{Machleidt:2000ge}, respectively. We use the on-shell binary amplitudes and a monopole form factor with the realistic value of $\beta$ in Eq.~(\ref{mb3}) for the off-shell correction to the $S$-wave $NN\!\to NN$ amplitude. According to Nakamura \textit{et al.}~\cite{Nakamura:2018cst} (Fig.~6 there) the off-shell effects visibly reduce the calculated \crss\ for $\gdpi$ at $\ega~\sim~300$~MeV and are rather small at higher energies $\ega~\sim~700$~MeV. Also these effects as the FSI ones are more pronounced in the $\gdpipn$ channel. \\
(ii) Secondly: we transformed the proper relation between the $\gN$ and $\gd$ \diff\ \crss\, i.e., Eq.~(\ref{ne2}), to simplified Eqs.~(\ref{ne7}) and (\ref{ne8}) where the $\gNpiN$ \crs\ is averaged over the photon energy (or effective $\pi N$ mass $W$) around the $\ega$ value. This approximation was used in practice when extracting the $\gnpiN$ \diff\
\crss\ from the deuteron data. Comparisons of the $\gnpiN$ \diff\ \crss\ obtained from the model-generated $\gd$ ones using the proper and simplified extraction formulas are given in Ref.~\cite{Nakamura:2018fad}. The difference in the results (Fig.~1 there) is rather significant (especially for the $\gnpin$ channel) at $\ega~=~300$~MeV ($\Delta(1232)$ region), where the $\gnpiN$ amplitude changes rapidly with energy. At high energy $\ega~=~1000$~MeV the simplified extraction procedure reproduces well the results of the proper one due to a weak energy behavior of the $\gnpiN$ except the forward region $\cos\theta > 0.5$ where a small difference remains. We accept the critique on both points discussed. However, we expect that the approximations mentioned here (on-shell binary amplitudes and simplified extraction formula) do not essentially affect the analysis of high-energy $\gnpip$ data in our papers~\cite{Chen:2012yv,CLAS:2017dco}. The $\gnpin$ \diff\ \crss\ from the Mainz A2 experiment on the $\gdpipn$ reaction~\cite{A2:2019yud} was extracted using the FSI-correction factor $R$ in Eq.~(\ref{r3}) that accurately accounts for kinematics. In this case the obtained $\gnpin$ were not averaged over the invariant mass $W$. \\
(iii) Thirdly: Nakamura~\cite{Nakamura:2018cst,Nakamura:2018fad} applied exactly the kinematical cuts used in the experimental analyses~\cite{CLAS:2017kua,CLAS:2017dco,Briscoe:2012ni} of the $\gdpipp$ data. We use more simple and slightly different condition, given in Eq.~(\ref{cut}). We suppose to use more precise kinematical cuts in future analyses. However, extracting~\cite{A2:2019yud} the $\gnpin$ \crss\ we use the factor $R$ from Eq.~(\ref{r3}) and automatically include all kinematical constraints on the $\gdpipn$ events. 

\vspace{3mm}
\section{Neutron Couplings}
\label{Sec:Couplings}
\vspace{2mm}

The SAID MA19 solution provides new results on the helicity amplitudes $A_{1/2}$ and $A_{3/2}$ of the photon decays $N^\ast\to\gamma n$ for the $N(1440)1/2^+$, $N(1520)3/2^-$, and $N(1535)1/2^-$ and $N(1650)1/2^-$ states (Table~\ref{tab:pole}). Here comparisons are made with the Bonn-Gatchina (BnGa)~\cite{Anisovich:2017xqg,Anisovich:2013jya} values and with an earlier SAID determination. For the $N(1520)$, the PDG2018~\cite{ParticleDataGroup:2018ovx} lists only Breit-Wigner (BW) values. This being the first determination of pole values, we compared at the level of moduli, finding good agreement. The agreement between BW and pole values is not as good for the Roper resonance, where the complicated pole-cut structure may invalidate this simple comparison of pole and BW quantities.
\begin{table}[th]
\caption{Moduli [in (GeV)$^{-1/2}\times 10^{-3}$] and phases (in degrees) of the photon-decay amplitudes $N^\ast \to n \gamma$ at the pole for $A_{1/2}$ and $A_{3/2}$ from the SAID MA27~\cite{Mattione:2011} and MA19~\cite{A2:2019yud} solutions. Pole results from the Bonn-Gatchina (BnGa) analysis are included for comparison~\cite{Anisovich:2017xqg} (BW values are from Ref.~\cite{Anisovich:2013jya}). BW values labeled with $^\dagger$.
}
\begin{center}
\centering
\small\begin{tabular}{|c|ccc|}
\hline\rule{0pt}{13pt}
 Resonance/Coupling&         ~SAID MA19~             & ~SAID MA27~                   & ~BnGA~               \\
\hline\rule{0pt}{13pt}
$N(1440)1/2^+$~$A_{1/2}$&  80$\pm$10,~~$( 96\pm  2)^\circ$& 65$\pm$5,~~$(  5\pm  3)^\circ$&  43$\pm$12$^\dagger$ \\
$N(1520)3/2^-$~$A_{3/2}$&-130$\pm$ 8,~~$( 20\pm  6)^\circ$&                               &-113$\pm$ 2$^\dagger$ \\
~~~~~~~~~~~~~~~~~~~$A_{1/2}$& -47$\pm$ 4,~~$(  1\pm  2)^\circ$&                               & -49$\pm$ 8$^\dagger$ \\
$N(1535)1/2^-$~$A_{1/2}$& -70$\pm$10,~~$(  2\pm  5)^\circ$&-55$\pm$5,~~$(  5\pm  2)^\circ$& -88$\pm$ 4,~~~$(5\pm 4)^\circ$  \\
$N(1650)1/2^-$~$A_{1/2}$&  13$\pm$ 4,~~$(-50\pm 15)^\circ$& 14$\pm$2,~~$(-30\pm 10)^\circ$&  16$\pm$ 4,~~~$(-28\pm 10)^\circ$ \\
\hline
\end{tabular}
\end{center}
\label{tab:pole}
\end{table}

\vspace{3mm}
\section{The Reactions $\gdpiNN$ in the Near-Threshold Region}
\label{Sec:Thr}
\vspace{2mm}

The pion \photo\ processes on a deuteron are interesting from various points of view. Here are the study of a few-body interactions, extraction of the elementary amplitude and \crs\ of the $\gnpip$ reaction, testing various theoretical models, in particular, ChPT approach. The motivations for these investigations are also mentioned in the Introduction.

{\bf The reaction $\bfpipp$}.
Recently, the PIONS@MAX-lab Collaboration (MAX-lab, Lund, Sweden) has reported data on the total \crss\
of incoherent charged-pion \photo\ $\gdpipp$ close to threshold~\cite{Strandberg:2018djk,Strandberg:2017}. This study is focused on a determination of the total $\gnpip$ \crss\ on a ``neutron'' target, utilizing the deuteron
measurements, where model-dependent nuclear (FSI) corrections play a critical role.
The model, used here for analysing the deuteron $\gdpipp$ data, includes the following ingredients and approximations.
\vspace{1mm}

\begin{figure}\begin{center}
\includegraphics[width=4.45cm, keepaspectratio]{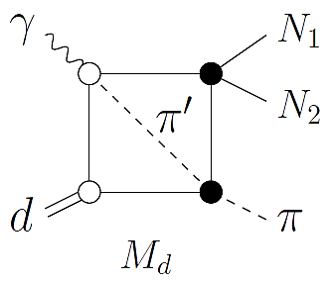}
\end{center}

\caption{Two-loop Feynman diagrams of the reaction $\gdpiNN$. The lines and blobs mean the same as in Fig.~\ref{diag}.}
\label{diag1}
\end{figure}

(1) We add the two-loop diagram $M_d$ in Fig.~\ref{diag1} to the terms, shown in Fig.~\ref{diag}. In the numerical calculations of the FSI amplitudes for the reaction $\gdpipp$ in Section~\ref{Sec:Reaction}, we neglected this diagram to save computer time.
\vspace{1mm}

(2) In the threshold region, we use the $S$-wave $\gnpip$ amplitude, approximated by the $E_{0+}$ multipole, taken to be constant. We include only the charged intermediate pion $\pi^-$ in the diagrams $M_c$ and $M_d$ since the contribution of intermediate $\pi^0$ is suppressed due to its small \photo\ $\gN\!\to\pi^0 N$ amplitudes. Thus, $\mgd\sim E_{0+}$
(hereafter $E_{0+}=E_{0+}(\gnpip)$). In this approximation,
\be
    \sigma(\gnpip)=4\pi\frac{k}{q_{\gn}}(E_{0+})^2\sim (E_{0+})^2,
    ~~~~\sigma(\gdpipp)=(E_{0+})^2\sigma_0.
\label{sgm}
\ee
Here: $q_{\gn}$ ($k$) is the CM momentum of the initial photon (final pion) in the reaction $\gamma n\!\to\!\pi^- p$; $\sigma_0$ is $\sigma(\gdpipp)$, calculated according to Eq.~(\ref{sigma}) with the factor $E_{0+}$ taken out of the amplitude $\mgd$, i.e., $\sigma_0$ does not depend on $E_{0+}$. The phase-space element $d\tau_3$ of the final $\pi^-pp$ system, used to calculate $\sigma_0$ in Eq.~(\ref{sgm}) through Eq.~(\ref{sigma}), can be written as
\be\ba{l}\dist
	d\tau_3=I\frac{Qp\,dwdzdz_1d\varphi_1}{2\pi(4\pi)^3\sqrt{s}},~~
    p=\sqrt{2\bar\mu w},
    \\ \rptt \dist
    Q=\sqrt{2\bar m(E^\ast-w)},~~~w=M_{pp}\!-2m_p.
\ea
\label{2}
\ee
Here: $I=1/2$ is identical factor (two protons); $E^\ast=\sqrt{s}\!-\mu-2m_p$ is the excess energy; $\mu$($m$) is the
$\pi^-$(proton) mass; $m=(m_p + m_n)/2$; $M_{pp}$ is the effective mass of the $pp$ system; $\bar m=2m\mu/(2m+\mu)$,
$\bar\mu=m\mu/(m+\mu)$; $z\!=\cos\theta$, $z_1\!=\cos\theta_1$; $\theta$ is the $\pi^-$ polar angle in the reaction rest frame; $\theta_1$ and $\varphi_1$ are the polar and azimuthal angles of relative motion in the $pp$ system. All the kinematical variables, needed to calculate the amplitude $\overline{|M_{\gd}|^2}$, can be expressed through $E^\ast$, $w$,
$z$, $z_1$ and $\varphi_1$.
\vspace{1mm}

(3) In the $NN$--FSI ($M_b$) and 2--loop ($M_d$) terms, the $S$-wave $pp$-scattering amplitude, which also includes the Coulomb effects, was taken from Ref.~\cite{Landau:1977}. We use the off-shell correction to the $pp$ amplitude according to prescription, given by Eq.~(\ref{mb3}).
\vspace{1mm}

(4) In the $\pi N$--FSI ($M_c$) and 2--loop ($M_d$) terms, the $S$-wave $\pi^- p$--scattering amplitude $\alpha_{\pi^-p}=\beta_0-\beta_1$ is used with the isospin scattering lengths ($\beta_0, \beta_1)=(-28, -881)$ in $10^{-4}/m_{\pi}$ units~\cite{Doring:2004kt}. The quoted determination of the pion-nucleon coupling constants is in good agreement with later determination~\cite{Baru:2011bw}, where $\alpha_{\pi^-p}=(85.66\pm 0.14)\times 10^{-3}/m_{\pi}$.
\vspace{1mm}

(5) The DWF of the Bonn potential is used in the parametrization from Ref.~\cite{Machleidt:2000ge}. Both $S$- and $D$-wave components of DWF are included in the IA diagram $M_a$, while $D$-wave is neglected in the diagrams $M_b$, $M_c$, and $M_d$.

As is seen below (Fig.~\ref{pipp-th}), the contributions of the terms $M_c$ and $M_d$, containing $\pi N$ rescattering, are relatively small in comparison with $M_a$ and $M_b$, due to a small value of the $\pi N$-scattering length. The 2-loop diagram which can be obtained from the $NN$-FSI term $M_b$, when adding the final $\pi N$ rescattering, is suppressed by factor $m_{\pi}/m_N$ compared to the 2-loop term $M_d$ and here is neglected.

Let us estimate missing contributions, including other intermediate mesons in the diagrams giving $M_c$ and $M_d$. An example is the $\eta$ meson. The $\pi N$ and $\eta N$ channels have a large coupling due to the $N(1535)$ resonance. Consider the diagram (denoted by $M^{\eta}_c$), obtained from the $\pi N$-FSI term $M_c$, when the intermediate pion is replaced by an eta-meson. This diagram will include the amplitudes $E_{0+}(\eta p)\equiv E_{0+}(\gamma p\to\eta p)$ and $a_{\eta n\to\pi^-p}$. Compare the products $E_{0+}(\eta p)\times a_{\eta n\to\pi^-p}$ and $E_{0+}(\gamma n\to\pi^-p)\times a_{\pi^-p}$, where $a's$ are the scattering lengths. Estimating $E_{0+}(\eta p)$ and $a_{\eta n\to\pi^-p}$ through the $s$-channel N(1535) mechanism at the resonance energy and making use of the known N(1535)-decay parameters, we obtain this product in the $\eta$ case to be $\sim~1.5$ times smaller than in the $\pi$ case. Also, the loop integral in the eta case should be suppressed compared to the pion case, since at the $\pi^-pp$ threshold we are essentially below the $\eta pn$ threshold in the intermediate state; thus, we neglect the $M^{\eta}_c$ term. If one replaces the intermediate pion by eta in the 2-loop term $M_d$ the arguments are the similar.

The amplitude $\mgd$ on the deuteron can be performed as
\be\ba{c}
    \mgd = c\varphi^+_1(L + i\bfK\cdot\bfsig)\varphi^c_2,~~~
    c = 16\pi W\sqrt{m}~~ (W = m + \mu),
    \\ \rpt
    L = L_a + L_b+L_c + L_d,~~~~~~ L_a = L^{(s)}_a + L^{(d)}_a,
    \\ \rpt
    \bfK\! = \!\bfK_a\! + \!\bfK_b\! + \!\bfK_c\! + \!\bfK_d,~~~~~
    \bfK_a\! = \!\bfK^{(s)}_a\! + \!\bfK^{(d)}_a.
\ea
\label{th1}\ee
Here, the subscripts $``a,b,c,d''$ mean the contribution of the amplitudes $M_{a,b,c,d}$; the superscripts $``(s),(d)''$ --
the contributions from $S$- and $D$-wave components of DWF.
With the above-listed assumptions the terms $L_{a,b,c,d}$ and $\bfK_{a,b,c,d}$ can be written as follows.\\
(a)~\underline{IA terms}:
\be\ba{c}
    L^{(s)}_a = x_aE_{0+}(\bfe\cdot\bfeps),~~~ x_a\! = \!f_1\! + \!f_2,
    \\ \rptt
    L^{(d)}_a = -\Bigl[g_1(\bfe\cdot\bfn_1)(\bfeps\cdot\bfn_2) 
    + g_2(\bfe\cdot\bfn_2)(\bfeps\cdot\bfn_1)\Bigr]E_{0+},
    \\ \rpt
    \bfK^{(s)}_a = y_aE_{0+}[\bfe\times\!\bfeps],~~~ y_a\! = \!f_2\!-\!f_1,
    \\ \rptt
    \bfK^{(d)}_a = \Bigl(g_2(\bfeps\cdot\bfn_2)[\bfn_2\!\times\!\bfe] 
    - g_1(\bfeps\cdot\bfn_1)[\bfn_1\!\!\times\!\bfe]\Bigr)E_{0+};
    \\ \rule{0pt}{22pt}\dist
    f_{1,2}\! = \frac{u(p_{1,2})}{\sqrt{2}} + \frac{w(p_{1,2})}{2},
    ~~~g_{1,2}\! = \frac{3}{2}\,w(p_{1,2}),~~~\bfn_{1,2}\! = \frac{\bfp_{1,2}}{p_{1,2}}.
\ea
\label{th2}\ee
Here, $\bfn_{1,2}$ -- the unit vectors; $u(p)$ and $w(p)$ are the $S$- and $D$-wave parts of the DWF~\cite{Machleidt:2000ge}. We use normalization $\int\!\!d\bfp\,\,[u^2(p)+w^2(p)]=(2\pi)^3$.
\\
(b)~\underline{$NN$-FSI terms}:
\be\ba{c}
    L_b = x_b E_{0+}(\bfe\cdot\bfeps),~~~ \bfK_b = 0,
    ~~~x_b = 2 I^{}_{N\!N}f^{}_{N\!N}(p),
\\ \rptt\dist
    I^{}_{N\!N}\! = I(p^2,\Delta) - I(-\beta^2,\Delta),~~~~\Delta = |\bfdeL|,
\\ \rttt\dist
    I(a^2,\Delta)\equiv\!\int\!\frac{d\bfx\,u(|\bfx\! + \!\bfdeL|)}
    {2\pi^2\sqrt{2}\,(x^2\!-\!p^2\!-i0)},
    ~~~\bfdeL = \frac{1}{2}(\bfp_1\!+\!\bfp_2).
\ea
\label{th3}\ee
Here: $f^{}_{N\!N}(p)$ is the on-shell $S$-wave $pp$-scattering amplitude with Coulomb corrections~\cite{Landau:1977}; $\beta$ is the off-shell parameter, given in Eq.~(\ref{mb3});
The integral $I(a^2\!,\Delta)$ is written out in Eq.~(A.10) of Ref.~\cite{Briscoe:2020qat}.
\\
{c)~\underline{$\pi N$-FSI terms}:
\be\ba{c}
    L_c\! = x_cE_{0+}a_{\pi N}\,(\bfe\cdot\bfeps),
~~~~x_c\! = I_1\! + \!I_2; 
\\ \rptt\dist
    \bfK_c\! = y_cE_{0+}a_{\pi N}\,[\bfe\!\times\!\bfeps],
~~~~y_c\! = I_1\!-\!I_2;
\\ \rttt\dist
    I_i\! = \!I(k^2_i,\Delta_i),~~\Delta_i = |\bfdeL_i|,
~~\bfdeL_i\! = \!\frac{m}{m\! + \!\mu}(\bfk\! + \!\bfp_i).
\ea
\label{th4}\ee
Here: $k_i$ are the relative momenta in the pion-proton pairs $\pi^-p_i$ ($i\!=\!1,\!2$); $a_{\pi N}$ is the $\pi^-p\,$-scattering
length.
\\
{d)~\underline{2-loop terms}:
\be\ba{c}
    L_d = x_d E_{0+}(\bfe\cdot\bfeps),~~~~ \bfK_d = 0,
~~~x_d = 2K(p,b,\Delta)f_{N\!N}(p)a_{\pi N},
\\ \rtttt\dist 
    K(p,b,\Delta) = \frac{m\! + \!\mu\!}{m}\int\!\! 
    \frac{d\bfx d\bfy\,u(|\bfx\! + \!\bfy\! - \!\Delta|)f(x,p)}
    {4\pi^4\sqrt{2}\,\,(x^2\! - \!p^2\! - i0)(y^2\! - \!b^2\!-i0)},
\\ \rule{0pt}{20pt}\dist
    \Delta = |\bfdeL|,~~~\bfdeL\! = \frac{1}{2}(\bfq\! + \!\bfk),~~~
    b^2\! = \!2\mu(\sqrt{s} - \sqrt{s_0})\ge 0.
\ea
\label{th5}\ee
Here, $\sqrt{s_0}\!=2m^{}_N\!+\!\mu$; $f(x,p)$ is given in Eq.~(\ref{mb3}); the denominator $(y^2\!-\!b^2\!-\!i0)$ of the pion propagator is obtained, neglecting the kinetic energies (static approximation) of the intermediate nucleons. The expression for $K(p,b,\Delta)$ is given in Eqs.~(A.11) and (A.12) of Ref.~\cite{Briscoe:2020qat}.
\begin{figure}\begin{center}
\includegraphics[width=10cm, angle=90, keepaspectratio]{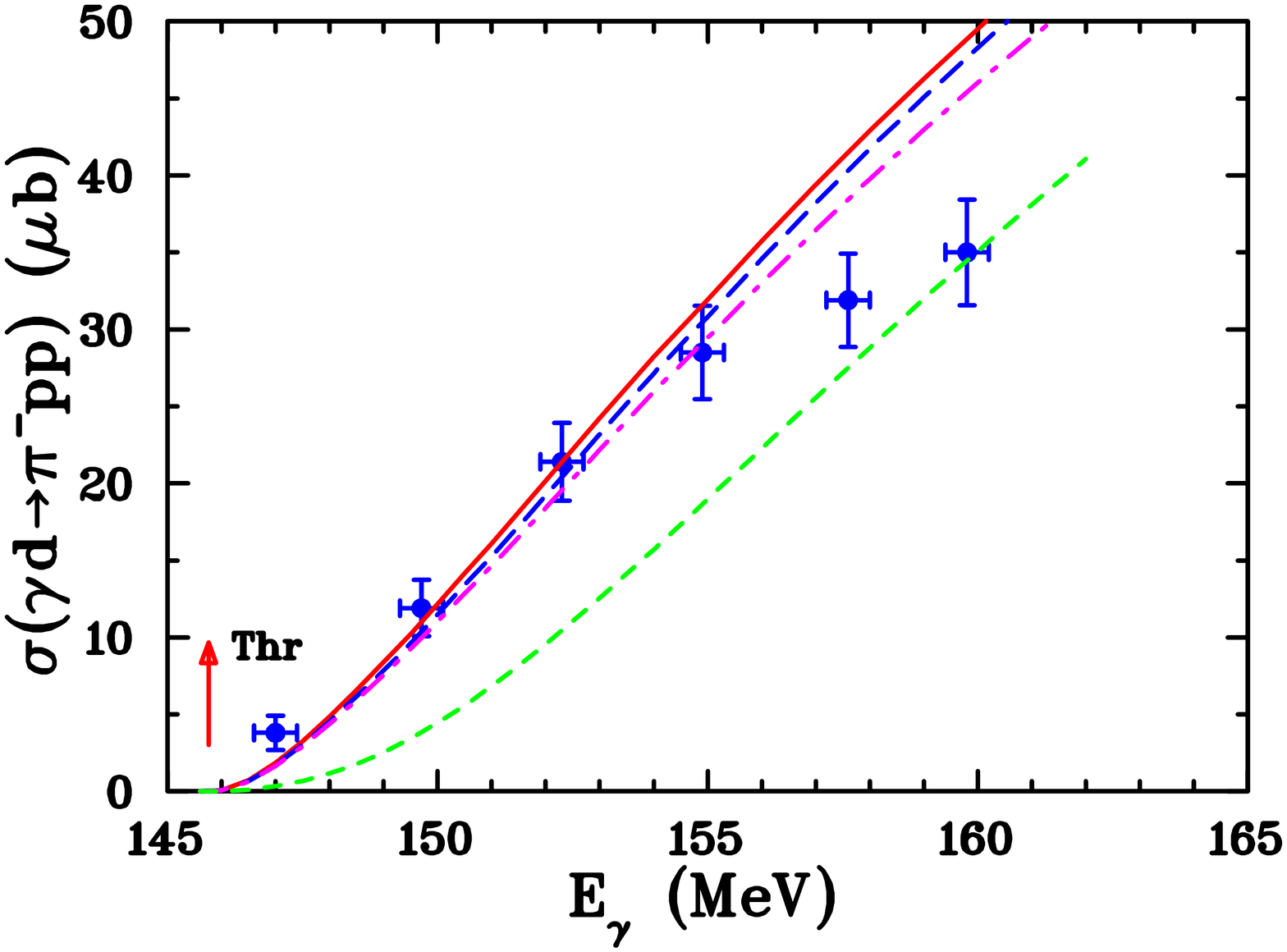}
\end{center}
\vspace{-1cm}

\caption{Total \crss\ of the reaction $\gdpipp$: blue filled circles are the  MAX-lab data~\cite{Strandberg:2018djk,Strandberg:2017}; $\ega$ is the photon energy in the laboratory frame; the statistical and systematic uncertainties from Table~II of Ref.~\cite{Strandberg:2018djk,Strandberg:2017} are summed in quadrature. The green short dotted curve shows the result, obtained with the IA amplitude $M_a$. Successive addition of $M_b\,$($NN$-FSI), $M_c\,$($\pi N$-FSI) and $M_d\,$(2-loop) terms in Fig.~\ref{diag1} leads to the magenta dash-dotted, blue long dashed, and red solid curves, respectively.}
\label{pipp-th}
\end{figure}

The square of the amplitude $\mgd$~(\ref{th1}) is $|\mgd|^2= 2c^2 (|L|^2+|\bfK|^2)$ for unpolarized nucleons. Averaged over the photon and deuteron polarization states it reads
\be
    \overline{|\mgd|^2} = 2c^2 (\overline{|L|^2}+\overline{|\bfK|^2}).
\label{th6}\ee
The expressions for $\overline{|L|^2}$ and $\overline{|\bfK|^2}$ obtained, making use of Eqs.~(\ref{th1})--(\ref{th5}), are given in Eqs.~(A.8) of Ref.~\cite{Briscoe:2020qat,Strandberg:2017}.

The total cross sections of the reaction $\gdpipp$, measured by PIONS@MAX-lab
Collaboration~\cite{Strandberg:2018djk,Strandberg:2017} 
are shown in Fig.~\ref{pipp-th}. Fitting these data by Eq.~(\ref{sgm}) for $\sigma(\gdpipp)$, using $E_{0+}$ as a free parameter, we obtain $E_{0+}(1-6)=-31.86\pm 0.8$ (in $10^{-3}/m_{\pi}$ units). The notation $(1-6)$ means that the $\chi^2$ fit includes all six data points in Fig.~\ref{pipp-th}. The solid curve shows the \crss, calculated with the total amplitude $\mgd$. The other curves are as explained in the figure caption. One can see that the main effect of FSI comes from the $N\!N$-FSI term $M_b$ (compare dash-dotted and long dashed curves in Fig.~\ref{pipp-th}), while the contributions of the terms $M_c$ and $M_d$ are small.

A relatively large disagreement of the model with the data is observed close to threshold at $\ega~=~147$~MeV.
Excluding this 1st data point from the fit, we obtain $E_{0+}(26) = -31.75\pm 0.8$ (in the same units). Both
variants, (16) and (26), are in agreement with the values $E_{0+}=-32.7\pm 0.6$, $-31.9$, and $-31.7$ from
Refs.~\cite{Bernard:1996ti} (ChPT), \cite{Drechsel:1992pn} (Born terms in pseudovector coupling), and
\cite{Hanstein:1996bd} (dispersion relations), respectively. The model also overestimates the data above $\ega~\sim~156$~MeV. If one excludes the 5th and the 6th data points at $\ega~=~157.6$~MeV and 159.8~MeV in
Fig.~\ref{pipp-th}, then the $\chi^2$ fit gives $E_{0+}(14) = -33.70\pm 1.2$ and $E_{0+}(24) = -33.94\pm 1.2$.

Suppose this discrepancy comes from the model approximations with energy-independent $\gnpip$ amplitude $E_{0+}$~=~constant. Let us briefly discuss the effects, not included here, connected with the energy dependence of the $E_{0+}$ and $P$-wave contributions to the $\gnpip$ amplitude. We can roughly estimate these corrections from the results of Ref.~\cite{Lensky:2005hb} on the reaction $\gdpinn$ in the chiral perturbation theory, where the Born $\gnpip$ amplitudes (with a Kroll-Ruderman term) in the threshold region were used. At $\Delta\ega = \ega-E_{th} = 15$~MeV ($E_{th}$ is the threshold energy), the energy-dependent correction to the constant $E_{0+}$ decreases the total \crss\ by $\sim 6\%$ (Fig.~8 of Ref.~\cite{Lensky:2005hb}), while the $P$-wave contribution increases it by $\sim 3\%$ (Fig.~9 there). Approximately the same corrections for $\gdpipp$ seem to be not enough to account for the discrepancy in Fig.~\ref{pipp-th} above $\ega~\sim~156$~MeV. Finally, in view of the visible disagreement of the model with deuteron \crss\ in the last two energy points, we cannot prefer any alternative among the fits given above.
Our fitted values of $E_{0+}$, within errors, and the results of earlier estimates are mostly overlapping. To obtain a more precise value of $E_{0+}$ from the threshold deuteron data, one should improve the model, especially to improve the theoretical description of the last two energy points of the deuteron \crss. We leave these details for a future study.
\begin{table}[t]
\caption{Total \crs\ for $\pi^-$ \photo\ on the neutron with statistical and systematic uncertainties in quadrature (4th, 5th, and 6th columns present partial components of statistical and systematic uncertainties).
}\vspace{3mm}
\centering
\begin{tabular}{|ccccc|}
\hline\rule{0pt}{13pt}
  $\ega$         &    $\sigma$   & ~Exp~  & ~Exp~  & $E_{0+}$~fit \\
                 &               & ~Stat~ & ~Sys~  & ~Sys~ \\
  (MeV)          &    ($\mu$b)   &  (\%)  &  (\%)  &  (\%)\\
\hline\rule{0pt}{13pt}
  149.7$\pm$ 0.4 & ~31.9$\pm$ 9.5 &  5.3   &  28.9 &  4.9 \\
  152.4$\pm$ 0.4 & ~56.0$\pm$ 9.0 &  2.5   &  15.1 &  4.9 \\
  155.0$\pm$ 0.4 & ~71.2$\pm$ 9.1 &  1.4   &  11.7 &  4.9 \\
  157.6$\pm$ 0.4 & ~83.1$\pm$ 9.8 &  1.8   &  10.5 &  4.9 \\
  160.3$\pm$ 0.4 & ~93.4$\pm$10.0 &  1.3   &   9.4 &  4.9 \\
  162.5$\pm$ 0.4 & 100.7$\pm$11.0 &  1.4   &   9.7 &  4.9 \\
\hline
\end{tabular}
\label{tab:gnpip}
\end{table}

Table~\ref{tab:gnpip} shows the \crss\ $\sigma(\gnpip)$ from Eq.~(\ref{sgm}) at $E_{0+}=E_{0+}(1-6)=-31.86\pm 0.8$. The results are given at the same values $\Delta\ega = \ega - E_{th}$ as in Fig.~\ref{pipp-th}, i.e., the $\ega$'s are shifted by the difference ($148.44 - 145.76$)~MeV of the $\gnpip$ and $\gdpipp$ threshold energies. Total uncertainties include statistical and systematic uncertainties of the MAX-lab experimental data with the FSI contribution.

The extracted $\gnpip$ \crss\ from Table~\ref{tab:gnpip} are shown in Fig.~\ref{gnpip-th} with previous measurements~\cite{White:1960ukk,Salomon:1983xn,Liu1:994}. These data are in good agreement with predictions from previous phenomenological analysis, such as SAID and MAID.
\begin{figure}\begin{center}
\vspace{-2.7cm}
\includegraphics[width=10cm, angle=90, keepaspectratio]{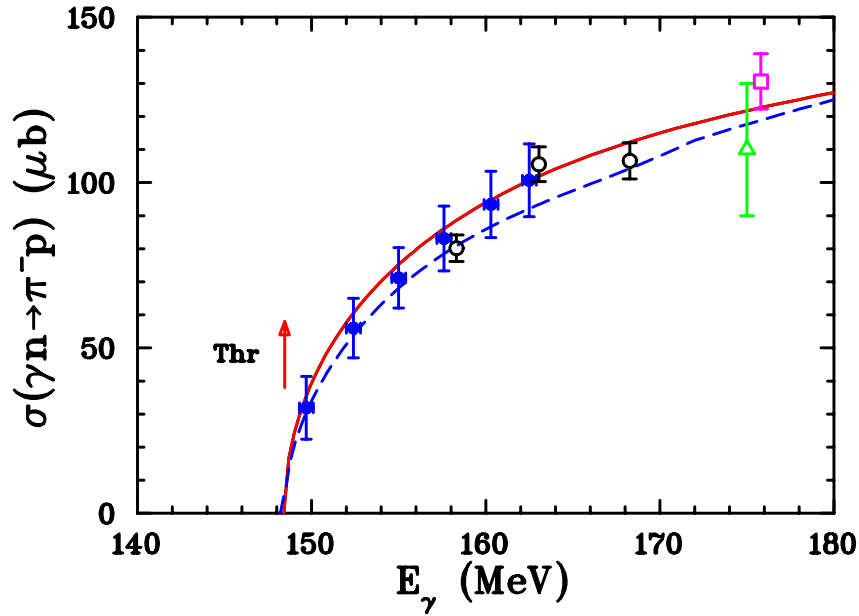}
\end{center}
\vspace{-1.2cm}

\caption{Total \crs\ of the reaction $\gnpip$. Previous measurements for the inverse reaction $\pi^-p\to\gamma n$ are from Cornell Synchrotron~\cite{White:1960ukk} (green open triangle), and TRIUMF~\cite{Salomon:1983xn} (magenta open square) and \cite{Liu1:994} (black open circles). Statistical and systematic uncertainties are summed in quadrature. Red solid (blue dashed) curves are predictions by the SAID MA19~\cite{A2:2019yud} (MAID2007~\cite{Drechsel:2007if}) solution.}
\label{gnpip-th}
\end{figure}

{\bf The reaction $\bfpinn$}.
Similar model calculations were made for the $\gdpinn$ total \crss\ in the near threshold region. The obvious changes for this case in Eqs.~(\ref{th2})--(\ref{th5}) mean that $E_{0+}$ is the $\gppin$ multipole, $f_{N\!N}$ is the on-shell $^1S_0$ $nn$-scattering amplitude, $a_{\pi N}$ is the $\pi^+n\,$-scattering length. Here we use the amplitude $f_{nn}$ in two variants
\be\ba{l}
    (a)~~~  f^{}_{nn}(p)=(-a^{-1}_{nn}+\half r^{}_{nn}p^2-ip)^{-1},
    ~~~a_{nn}=18.9~{\rm fm},~~~r_{nn}=2.75~{\rm fm};
    \\ \rpt
    (b)~~~ f^{}_{nn}(p)~~{\rm from~Ref.~[55]~(Appendix~B)}.
\ea
\label{th7}\ee
Variant (a) shows the on-shell amplitude. The off-shell dependence of $f_{nn}$ is introduced according to Eq.~(\ref{mb3}) with the same value of $\beta$. The $^1S_0$ amplitude $f_{nn}$ of variant~(b) was obtained in Appendix~B of Ref.~\cite{Lensky:2005hb} from a separable interactions which approximate the CD-Bonn potential~\cite{Machleidt:2000ge} end exactly reproduce its on- and off-shell properties. With $f_{nn}$ from variant~(b) Eqs.~(\ref{th3}) and (\ref{th5}) for $NN$-FSI and 2-loop terms take some more complicated forms omitted here.
Fig.~\ref{pinn-th} show the model predictions (curves) for the total $\gdpinn$ \crss\ in the near threshold region and the experimental data~\cite{Booth:1979zz}. The model results include variants (a) and (b) of $f_{nn}$ from Eq.~(\ref{th7}) and different input values of the $E_{0+}(\gppin)$ multipole. The plots: (a) corresponds to the value $E_{0+}=28.2$ (in $10^{-3}/m_{\pi}$ units) from ChPT~\cite{Bernard:1996ti}; (b) and (d) -- the results at $E_{0+}=27.7$~\cite{Drechsel:1992pn}; (c) -- the results at threshold value $E_{0+}=27.03$, obtained from our calculation of the Born $\gppin$ amplitude.  
Fig.~\ref{pinn-th} shows the best theoretical descriptions of the data, which came from the Bates Linear Accelerator, in the plots (b)--(c). Here our predictions are quite close to the results of Lensky \textit{et al.}~\cite{Lensky:2005hb}, obtained in ChPT.
An essential question arises when we compare the results for  $\gdpipp$ and $\gdpinn$ in Figs.~\ref{pipp-th} and \ref{pinn-th}. It concerns a different quality of theoretical description of the data on these channels.

When we compare the results for $\gdpipp$ and $\gdpinn$ in Figs.~\ref{pipp-th} and \ref{pinn-th} we see an essential difference in the quality of theoretical descriptions of the data in these channels. This raises two concerns. The first one is a possible disadvantage of the model in connection with description of the $\gdpipp$ data, discussed above. The second is focused on the quality of the $\gdpipp$ data in Fig.~\ref{pipp-th}. 
\begin{figure}\begin{center}
\includegraphics[width=12cm, keepaspectratio]{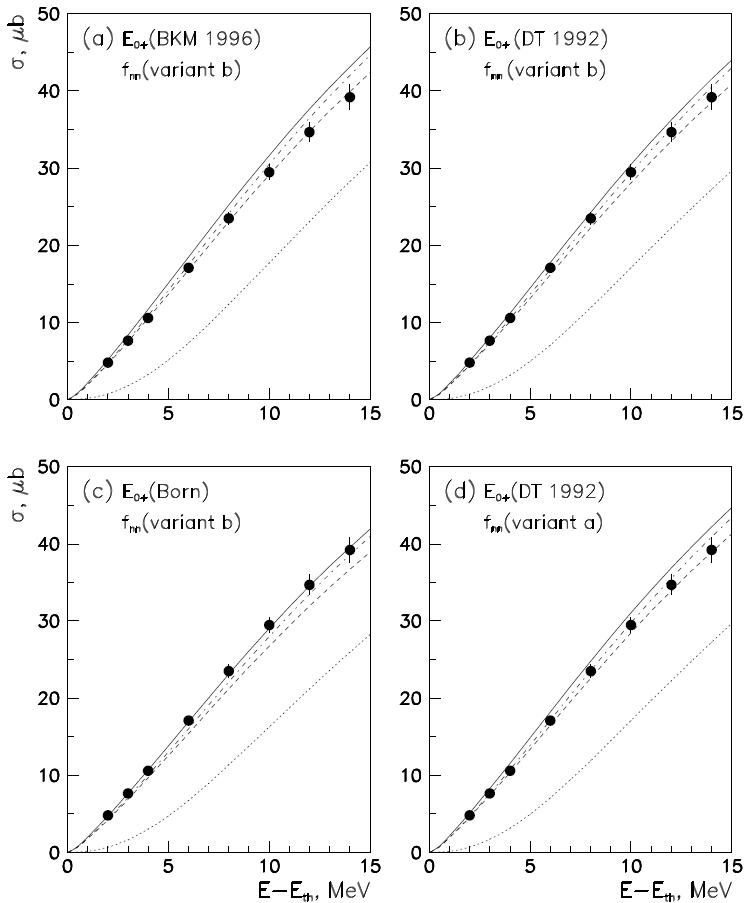}
\end{center}
\vspace{-7mm}

\caption{Total \crs\ of the reaction $\gdpinn$. Filled circles are the data from the Bates Linear Accelerator~\cite{Booth:1979zz}. The curves show the model predictions. The notations of curves are the same as in Fig.~\ref{pipp-th}. The plots correspond to different threshold values of the $\gppin$ multipole $E_{0+}$ or different variants of $^1S_0$ $nn$-scattering amplitude $f_{nn}$. The plots (a)--(c) are obtained with variant~(b) of Eq.~(\ref{th7}), the plot (d) -- with variant~(a). The values of $E_{0+}$ are given in the text.}
\label{pinn-th}
\end{figure}

\vspace{5mm}
\section{Conclusion}
\label{Sec:Concl}
\vspace{2mm}

Here we have presented a review and an expansion of our results concerning the theoretical description of pion \photo\ reactions and the extraction from deuteron data of the $\gnpiN$ \diff\ \crss\ used to determine the photodecay amplitudes $N^\ast\!\to\!\gamma n$ for the excited baryon $N^\ast$ states. The model amplitude involves impulse-approximation, $NN$- and $\pi N$-FSI terms. The input for these amplitudes includes the $\gN,\pi N\to\pi N$ and $NN\to NN$ binary amplitudes and the deuteron wave function taken from the known parametrizations mentioned above. A reasonable description of available data on the $\gdpipp$ \diff\ \crss\ over a wide range of photon energies and outgoing-pion angles was obtained. The model predictions (with $\pi N$-FSI term neglected) for the $\gdpipn$ \diff\ \crss\ notably overestimate the experimental values. Data on $\gdpipp$~(CLAS Collaboration) and $\gdpipn$~(A2 Collaboration at MAMI) were used to extract the $\gnpip$ and $\gnpin$ \crss\ $\dso^{cm}_{\pi}$, respectively. The obtained $\gnpip$ \crss\ rises more sharply at forward angles compared to previous measurements. This indicates a disadvantage of the model with rather strong $NN$-FSI effect at small angles (pion production angle less than 30$^\circ$ in c.m.). However, the influence of this feature on the extracted $\gnpip$ \crss\ is essentially suppressed due to kinematical cuts, excluding small angles $\theta^{cm}_{\pi}$, used in the $\gdpipp$ data.

Based on the PWA of the new $\gnpip,\pi^0n$ \diff\ \crss\ added to the world data, new results on the helicity amplitudes $A_{1/2}$ and $A_{3/2}$ of the photon decays $N^\ast\!\to\!\gamma n$ into a neutron channel for several excited baryon states above the $\Delta(1232)3/2^+$ region were reported~\cite{CLAS:2017dco,A2:2019yud}.

In several papers, our group has shown that FSI corrections on unpolarized measurements for reactions $\gamma d\to\pi NN$ are about 20\% (see, for instance, Refs.~\cite{Tarasov:2011ec,Tarasov:2015sta}). As polarization asymmetries measure ratios of cross sections, FSI effects are expected to have a considerably smaller effect on these and are expected to be comparable to, or less than, quoted systematic uncertainties from experimental sources. An indirect verification of this assumption is that the SAID group can fit new polarized measurements on a ``neutron'' target~\cite{DiSalvo:2009zz,Sokhan:2009,Graal:2010qrh,Dieterle:2017myg,CLAS:2017kua,A2atMAMI:2021iuz} when included with existing world data.

Predictions for the total \crss\ of charged-pion \photo\ reactions $\gd\to\pi^{\pm}NN$ near threshold are given in the model with a simplified $S$-wave $\gNpiN$ amplitude approximated by the $E_{0+}$ multipole, taken to be constant. The $\gdpipp$ \crss\ first measured by the PIONS@MAX-lab Collaboration were used to extract the $E_{0+}(\gnpip)$ multipole. The obtained $E_{0+}$ value is in reasonable agreement with the results of previous estimates. The model reasonably describes the near threshold $\gd\!\to\!\pi^-pp,\pi^+nn$ \crss\ with realistic values of $E_{0+}(\gnpip)$ and $E_{0+}(\gppin)$ except the region above $\ega~\sim~156$~MeV in the $\gdpipp$ case. This indicates a possible disadvantage of this  simplified model. However, in the $\gdpinn$ case the model description is better in the same excess-energy range. Thus, it would be desirable to have additional independent data on the $\gdpipp$ \crss\ near the threshold.

In spite of successful results, the model also demonstrates disadvantages with several of these issues mentioned above. Given the critical comments in Refs.~\cite{Nakamura:2018cst,Nakamura:2018fad} (Section~\ref{Sec:Analysis}), the model will be improved in future studies.

\begin{center}
{\large\bf Acknowledgements}
\end{center}
W.~J.~B. and I.~I.~S. were supported in part by the U.~S.~Department of Energy, Office of Science, Office of Nuclear Physics, under Award No.~DE–-SC0016583 and R.~L.~W. under Award No.~DE--SC0016582. 



\end{document}